\documentclass[fleqn,usenatbib]{mnras}

\usepackage{newtxtext,newtxmath}
\usepackage[T1]{fontenc}
\DeclareRobustCommand{\VAN}[3]{#2}
\let\VANthebibliography\thebibliography
\def\thebibliography{\DeclareRobustCommand{\VAN}[3]{##3}\VANthebibliography}

\usepackage{graphicx}	% Including figure files
\usepackage{adjustbox}  % For the orcid ids
\usepackage{amsmath}	% Advanced maths commands
\usepackage{wasysym}    % Extra astronomy symbols
\usepackage{bm}         % Bold maths
\usepackage{xspace}
\usepackage{caption}
\usepackage{enumitem}
\usepackage{cleveref}
\usepackage{dsfont}
\usepackage{multirow}

\makeatletter % Workaround to only color the year for cite commands
  \patchcmd{\NAT@citex}
    {\@citea\NAT@hyper@{%
      \NAT@nmfmt{\NAT@nm}%
      \hyper@natlinkbreak{\NAT@aysep\NAT@spacechar}{\@citeb\@extra@b@citeb}%
      \NAT@date}}
    {\@citea\NAT@nmfmt{\NAT@nm}%
    \NAT@aysep\NAT@spacechar\NAT@hyper@{\NAT@date}}{}{}

  \patchcmd{\NAT@citex}
    {\@citea\NAT@hyper@{%
      \NAT@nmfmt{\NAT@nm}%
      \hyper@natlinkbreak{\NAT@spacechar\NAT@@open\if*#1*\else#1\NAT@spacechar\fi}%
        {\@citeb\@extra@b@citeb}%
      \NAT@date}}
    {\@citea\NAT@nmfmt{\NAT@nm}%
    \NAT@spacechar\NAT@@open\if*#1*\else#1\NAT@spacechar\fi\NAT@hyper@{\NAT@date}}
    {}{}
\makeatother

\newcommand\Msun{\text{M}_{\astrosun}} % requires the wasysym package
\newcommand\HI{\ion{H}{I}\xspace} % neutral hydrogen
\newcommand\HII{\ion{H}{II}\xspace} % ionized hydrogen
\newcommand\CII{\ion{C}{II}\xspace} %
\newcommand\arepo{\mbox{\textsc{arepo}}\xspace}
\newcommand\areport{\mbox{\textsc{arepo-rt}}\xspace}
\newcommand\thesan{\mbox{\textsc{thesan}}\xspace}
\newcommand\thesanone{\mbox{\textsc{thesan-1}}\xspace}

\newcommand\orcid[1]{\href{http://orcid.org/#1}{\adjustbox{trim={-.15\width} {0\height} {-.15\width} {0\height},clip}{\includegraphics[height=10pt]{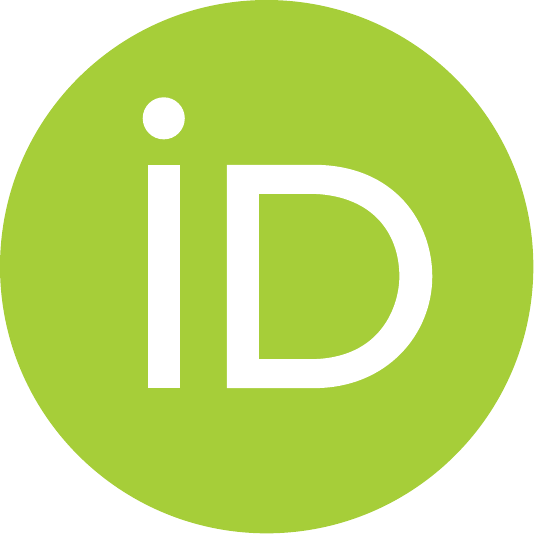}}}}

\title[Ly$\alpha$ intensity mapping in the EoR]{The \thesan project: Lyman-$\bmath\alpha$ intensity mapping of cosmic reionization}

\author[M.~Almualla et al.]{%
Mouza~Almualla\orcid{0000-0002-4694-7123},$^{1}$\thanks{E-mail: \href{mailto:mouza.almualla@cfa.harvard.edu}{mouza.almualla@cfa.harvard.edu}}
Aaron~Smith\orcid{0000-0002-2838-9033},$^{2}$
Rahul~Kannan\orcid{0000-0001-6092-2187},$^{3}$
Lars~Hernquist,$^{1}$
Enrico~Garaldi\orcid{0000-0002-6021-7020},$^{4}$
\newauthor
Adam~Lidz\orcid{0000-0002-3950-9598},$^{5}$
Kevin~Lorinc\orcid{0009-0005-3827-8774},$^{2}$
Jennifer~Yik~Ham~Chan\orcid{0000-0003-0314-7027},$^{6}$ and
Mark~Vogelsberger\orcid{0000-0001-8593-7692}$^{7}$
\\%
$^{1}$Center for Astrophysics $\vert$ Harvard $\&$ Smithsonian, 60 Garden Street, Cambridge, MA 02138, USA \\%
$^{2}$Department of Physics, The University of Texas at Dallas, Richardson, Texas 75080, USA \\%
$^{3}$Department of Physics and Astronomy, York University, 4700 Keele Street, Toronto, ON M3J 1P3, Canada \\%
$^{4}$ Kavli IPMU (WPI), UTIAS, The University of Tokyo, Kashiwa, Chiba 277-8583, Japan \\%
$^{5}$Department of Physics and Astronomy, University of Pennsylvania, 209 South 33rd Street, Philadelphia, PA 19104, USA \\
$^{6}$Department of Physics and Astronomy, Oberlin College, Oberlin, OH 44074, USA \\
$^{7}$Department of Physics $\&$ Kavli Institute for Astrophysics and Space Research, Massachusetts Institute of Technology, Cambridge, MA 02139, USA%
}

\date{Accepted XXX. Received YYY; in original form ZZZ}

\pubyear{2023}

\begin{document}
\label{firstpage}
\pagerange{\pageref{firstpage}--\pageref{lastpage}}
\maketitle

\begin{abstract}
Line Intensity Mapping (LIM) has garnered attention as a powerful cosmological probe, with next-generation instruments such as SPHEREx preparing to map the evolution of large-scale structure during the Epoch of Reionization (EoR). Lyman-alpha (Ly$\alpha$) emission in the EoR is strongly shaped by resonant absorption from neutral hydrogen in the diffuse intergalactic medium (IGM), which transforms galactic sources into a low surface-brightness background. In this work, we leverage the state-of-the-art \thesan cosmological simulations to produce high-resolution theoretical predictions for future Ly$\alpha$ LIM studies, constructing continuous light cones for line-of-sight cosmological integrations. We assess the contributions of recombination, collisional excitation, and unresolved \HII regions to the total Ly$\alpha$ spectral intensity. In addition, we explore the IGM in absorption at different redshifts using damping wing analysis. We produce channel maps exploring spatial fluctuations across redshift bands probe-able by LIM instruments. We find that the slope of the absorption-included Ly$\alpha$ fluctuation power spectrum at smaller scales ($k \gtrsim 2\times10^{-2}\,\rm{arcsec}^{-1}$) steepens toward lower redshift, and that our emission-only Ly$\alpha$ power spectrum lies above the SPHEREx sensitivity, whereas the absorption-included signal is $\sim$4 orders of magnitude lower--providing a conservative lower limit on inhomogeneity signatures and highlighting the importance of including resonant scattering in our model in the future. We also find that including outflows in a simple toy model boosts power by $4$ orders of magnitude. We identify limitations in our analysis and propose next steps, including incorporating the effects of resonant Ly$\alpha$ scattering and line interlopers, as well as larger simulation volumes.
\end{abstract}


%%%%%%%%%%%%%%%%%%%%%%%%%%%%%%%%%%%%%%%%%%%%%%%%%%

%%%%%%%%%%%%%%%%% BODY OF PAPER %%%%%%%%%%%%%%%%%%

\section{Introduction}
\label{sec:intro}

The Epoch of Reionization (EoR) marks a pivotal phase transition in cosmic history, central to numerous observational and theoretical efforts to better understand the evolution of our Universe. Following the recombination epoch at redshift $z \approx 1100$ culminating in the cosmic microwave background, the expanding Universe cooled sufficiently for free electrons and protons to bind as neutral atomic hydrogen, leading to a period known as the ``Dark Ages'' due to the absence of luminous sources. Within a few hundred million years after the Big Bang, the first stars and galaxies emerged from this darkness, emitting Lyman-Continuum (LyC) photons that began ionizing the previously neutral intergalactic medium \citep[IGM;][]{Dayal2018,Wise2019,GnMa2022,Robertson2022}.

Investigating the EoR is essential to understanding the formation and evolution of large-scale structure in the Universe. Despite significant advancements, there are still several critical open questions pertaining to when exactly reionization began and ended, and what the dominant sources driving this cosmic phase transition were. Mounting evidence has hinted towards a ``late'' reionization concluding around $z \sim 6$ \citep{Fan_2006,Mitra_2011,McMe2014}, possibly with a rapid progression at $z \sim 7\!-\!8$ \citep{NaiTacc2020}. However, more recent observations suggest that neutral regions may have extended to even lower redshifts \citep[e.g.][]{ChBe2021,ChBe2023}, while at the other extreme the first ionized bubbles were already in place at very early times \citep[e.g.][]{ZhuBe2024,Witstok2025}.

Advancing our knowledge of the EoR requires not only leveraging the wealth of data provided by dedicated ongoing and upcoming instruments, but also developing robust theoretical models and simulations for accurate forecasting and interpretation. In this context, a method known as Line Intensity Mapping \citep[LIM;][]{Kovetz2019,Bernal2022} provides a promising path towards probing the high-redshift Universe observationally \citep[e.g.][]{Visbal2010,Lidz2016,Liu2016,Breysse2017,Fonseca2017}. LIM targets specific atomic or molecular lines and, using low angular resolution, captures the integrated emission from all galaxies within each pixel of the telescope's field of view. By assuming that most of this emission originates from a single line, the spectral intensity can be translated into a redshift-space mapping, effectively creating a three-dimensional representation of the Universe over the range of redshifts targeted by the instrument. Ultimately, the LIM technique allows for the inclusion of faint galaxies that are otherwise undetectable individually, as it does not rely on resolving discrete objects.

While there has been special interest in mapping neutral hydrogen using the 21\,cm line, other spectral lines also offer valuable insights for LIM experiments. The commissioning of radio interferometers targeting 21\,cm signals include the Low-Frequency Array \citep[LOFAR][]{HaWi2013}, the Hydrogen Epoch of Reionization Array \citep[HERA;][]{DePa2017}\footnote{\url{https://reionization.org}}, and the forthcoming Square Kilometer Array \citep[SKA;][]{DeHa2009}, enhancing our understanding of the ionization and temperature structure of the IGM during reionization. Significant efforts have also been invested into the \CII fine structure line, which can be used as a tracer for both star formation and far-infrared dust emission \citep[e.g.][]{Gong_2012,SiSan2015,YueFe2015,Dumitru2019,KaMa2022}. Likewise, the rotational CO lines trace molecular clouds \citep[e.g.][]{RiHern2008,BrPa2014,SunCha2021,BeKo2022,Roy_2023}. Hydrogen lines such as H$\alpha$ and Ly$\alpha$ are well-established indicators of star-formation rates \citep{Pullen_2014,SiZa2017,CroMir2018}. In particular, the Ly$\alpha$ line, corresponding to the transition from the $n=2$ to $n=1$ energy levels in atomic hydrogen, provides pertinent information about both galaxies and the IGM during the EoR due to its resonant scattering properties, and is the focus of this paper.

Ly$\alpha$ photons exhibit relatively complex emission and transmission characteristics. They are produced primarily through recombinations in the interstellar medium (ISM) of high-redshift galaxies ionized by young, massive stars \citep{Dijkstra2014,KoBr2019,BeKo2022} and through collisional excitations, such as those resulting from shock heating and gravitational cooling radiation. Once generated, Ly$\alpha$ photons undergo multiple scatterings with neutral hydrogen atoms, escaping the ISM via a frequency and spatial diffusion process and transmitting through the IGM via Hubble flow redshifting further from line center. Ly$\alpha$ LIM can thus serve as an important probe of diffuse Ly$\alpha$ emission from the IGM, which is typically missed by conventional redshift surveys due to resonant scattering occurring far from the host source \citep{Fardal2001,Santos2004,Dijkstra2007,Faucher2010,Laursen2011,Zheng_2011,Gronke2020,Park2021,Smith2022}. This effectively converts Ly$\alpha$ emission to cosmic background light with a redshift-dependent correlation scale imprinted by the evolving intervening IGM. Additionally, Ly$\alpha$ LIM can aid in building the faint end of the Ly$\alpha$ luminosity function, as well as in untangling the features of the baryonic cosmic web \citep{WiPu2021,Byrohl2023,Liu2024,Tsai2024,Renard2024}.
% Generally, observing Ly$\alpha$ in absorption (through a background quasar's Ly$\alpha$ forest) and in emission (from  Ly$\alpha$ emitters) can help constrain the Ly$\alpha$ opacity--density relation \citep{ChBe2021}. 

It is no surprise, then, that upcoming instruments such as the Spectro-Photometer for the History of the Universe, Epoch of Reionization and Ices Explorer \citep[SPHEREx;][]{DoBo2014}, designed to perform an all-sky spectrophotometric survey across multiple bands, have identified Ly$\alpha$ as a key target line for high-redshift LIM, specifically aiming for the redshift range $5.2 < z < 8$. However, observations can be contaminated by emissions from other spectral lines at lower redshifts, referred to as \textit{foreground interlopers} \citep{CoYu2016}. SPHEREx is expected to have a high enough sensitivity to mitigate much of these contaminants through techniques like spectral masking. Moreover, performing cross-correlations between intensity maps of different lines at the same redshift provides a robust method for interloper removal.

Despite the potential of Ly$\alpha$ LIM, previous studies have largely relied on semi-analytical models or general purpose cosmological simulations. While the physical insights gained from these studies are crucial, ultimately these approaches may not fully capture the complexities of the EoR and the large-scale structures involved \citep[e.g.][]{Silva2013,Pullen_2014,Comaschi2016,ComaschiYueFerrara2016,MasRibas2017,MasRibas2020,MasRibas2023,Heneka2021}. In fact, due to the complexity Ly$\alpha$ LIM is often less developed than other spectral lines \citep{Moriwaki2018,Moriwaki2019,Dumitru2019,Padmanabhan2019,Sun2019,SunCha2021,Leung2020,Yang2021}. In this paper, we aim to provide more accurate predictions for Ly$\alpha$ LIM by utilizing the \thesan project \citep{Kannan2022,Smith2022,Garaldi2022,Garaldi2023}, a suite of state-of-the-art large-scale cosmological radiation-hydrodynamic simulations of the EoR. With a comoving box size of $L_\text{box} = 95.5$\,cMpc, these simulations offer a detailed and statistically robust framework for modeling Ly$\alpha$ emission and its interaction with the IGM, as a natural next step building upon the multi-tracer LIM study of \citet{kannan2022b}, as well as Ly$\alpha$ emitter explorations from \thesan \citep{Smith2022,Xu2023,Chen2025,Neyer2025}.

The \thesan simulations employ \textsc{arepo-rt} --- a moving mesh hydrodynamics code \citep{Springel2010} coupled with radiative transfer \citep{Kannan2019} --- to accurately model ionization fronts. They incorporate the IllustrisTNG model \citep{WeSp2017,PiNe2019,Springel2018} to simulate the physics of ionizing sources such as stars and active galactic nuclei, and include dust modelling as described in \cite{McTo2016,McTo2017}. By leveraging these simulations, we can capture the complex interplay between galaxies and the IGM during the EoR for more realistic predictions for upcoming LIM experiments like SPHEREx.

The paper is organized as follows: in Section~\ref{sec:methods}, we describe our methods, including a brief overview of the \thesan simulation suite and breaking down the pipeline in which we stitch Cartesian renders together to perform Ly$\alpha$ cosmological radiative transfer integrations along the line-of-sight (LoS); in Section~\ref{sec:results} we present our main findings, showing spectral intensity results including absorption only, both absorption and emission, as well as a rendering method to show the evolution of spatial fluctuations in Ly$\alpha$ intensity in a manner akin to LIM observations; finally, we discuss our results and conclusions in Section~\ref{sec:summary}.

\section{Methods}
\label{sec:methods}

\subsection{\thesan simulations}

As mentioned previously, the \thesan simulations are built off the IllustrisTNG galaxy formation model; more information on the details can be found in \citet{PiSp2017}, with the modifications introduced by \thesan outlined in \citet{Kannan2022}. Here, we provide a brief description of some aspects relevant to this work.

The IllustrisTNG and \thesan simulations both employ a Voronoi tesselation scheme to obtain a quasi-Lagrangian solution for the hydrodynamics at the interface of the mesh cells, and dark matter, gas, and photons are modeled self-consistently. Due to the resolutions involved, the ISM is treated on a sub-grid level as a two-phase gas; star-formation above the so-called equation of state threshold of $n_\text{H} \approx 0.1\,\text{cm}^{-3}$ is modeled stochastically using the Kennicutt-Schmidt relationship, while pressurization from supernovae is incorporated in the hot phase \citep{Springel2003}. AGN feedback is likewise implemented through sub-grid prescriptions, with ongoing improvements in its treatment in high-redshift simulations \citep{Bulichi_2025}. To account for cosmic variance, \thesan employs an approach to fix the amplitudes of the Fourier modes to the average power spectrum of the ensemble \citep{AnRa2016} rather than sampling from the Gaussian primordial power spectrum predicted by most inflationary models \citep{MuFe1992}

We base our analysis on the \thesanone simulation, the highest-resolution flagship run of the \thesan project. \thesanone is a radiation-magneto-hydrodynamic cosmological simulation with a large $(95.5\,\text{cMpc})^3$ volume evolved down to $z = 5.5$. \thesanone contains $2100^3$ dark matter and (initial) gas particles for resolutions of $m_\text{DM} = 3.12 \times 10^6\,\Msun$ and $m_\text{gas} = 5.82 \times 10^5\,\Msun$, effectively resolving atomic cooling haloes by at least 50 particles for a minimum cell size of $\sim10$ pc (physical) at $z=5.5$. \thesan is designed to self-consistently capture the formation and evolution of galaxies during the EoR and their feedback on the IGM. Most importantly, \thesan incorporates detailed galaxy physics and non-equilibrium thermochemistry with full on-the-fly radiative transfer and photo-ionization/heating of hydrogen and helium. We refer readers to \citet{Kannan2022} and \citet{Garaldi2023} for details on the simulation setup and public data release.

\thesan is built on the \areport code \citep{Kannan2019}, an extension of the moving-mesh \arepo code \citep{Springel2010, Weinberger2020}. The simulations couple the state-of-the-art IllustrisTNG galaxy formation physics model \citep{Weinberger2017, Pillepich2017, Springel2018}, an updated version of the original Illustris framework \citep{Vogelsberger2013, Vogelsberger2014b, Vogelsberger2014a}, to an on-the-fly radiative transfer module for ionizing photons across three energy bins over $[13.6, 24.6, 54.4, \infty)$\,eV, solved assuming the moment-based M1 closure relation \citep{Levermore1984} with second-order accuracy. A reduced speed of light approximation of $\tilde{c} = 0.2\,c$ is used for computational efficiency. The stellar sources are computed using the Binary Population and Spectral Synthesis models \citep[BPASS;][]{Eldridge2017, Stanway2018}, assuming a Chabrier initial mass function \citep[IMF;][]{Chabrier2003}. An additional birth cloud escape fraction parameter, $f_\text{esc}^\text{cloud} = 0.37$, is added to mimic the absorption of LyC photons below the resolution scale of the simulation, tuned to reproduce a realistic late-reionization history. Finally, gravity is calculated using a hybrid Tree--PM approach with updates and additional algorithmic details described in \citet{Springel2021}.

Our study complements others from the \thesan simulations, including emission line intensity mapping \citep{kannan2022b}, the impact of reionization on low-mass galaxies \citep{Borrow2023}, constraining galaxy populations \citep{kannan2023}, ionizing escape fractions \citep{Yeh2023}, alternative dark matter models \citep{Shen2024a}, galaxy sizes \citep{Shen2024b}, galaxy--IGM connections \citep{Garaldi2024a, Garaldi2024b, Kakiichi2025}, bubble size properties \citep{Neyer2024, Jamieson2025}, and Local Group reionization \citep{Zhao2025}.

\subsection{Redshift-space rendering fields}
\label{subsec:renderfields}
We now discuss various quantities obtained from the \thesan simulations. The simulation outputs contain a statistical rendering of the unstructured mesh onto a 3D uniform Cartesian grid that takes into account redshift-space distortions for important spectral quantities. As discussed, we are mainly interested in the Ly$\alpha$ emission and absorption properties, which probe the time-dependent Ly$\alpha$ radiation field from an ensemble of galaxies at once.

Peculiar motions of emitting and absorbing gas, on the order of $100\,\text{km\,s}^{-1}$, induce Doppler shifts relative to the Ly$\alpha$ line center in the cosmological observer frame. Therefore, the Ly$\alpha$ properties in the cartesian renders have been mapped onto a velocity-corrected spectral grid oriented along the positive $z$-axis. The observed redshift-space position, accounting for the effect of peculiar velocities, is given by the Hubble relation: $\Delta v = H(a) \Delta r = a H(a) \Delta r_c$, where the subscript denotes a comoving quantity. Note that $H(a)$ here is the Hubble parameter at scale factor $a$. Specifically, the adjusted comoving position for binning optically-thin emission is
\begin{equation}
  \bmath{s}_c = \bmath{r}_c + \frac{v_{\|}}{a H(a)} \hat{\bmath{r}}_{c,\|} \, ,
\end{equation}
where $v_{\|}$ is the LoS component of the peculiar velocity. During the EoR, it is valid to assume a matter-dominated Universe, allowing us to approximate the Hubble parameter as
\begin{equation}
  H(a) \approx H_0 \sqrt{\Omega_0} a^{-3/2} \, ,
\end{equation}
where $H_0$ is the present-day Hubble constant, and $\Omega_0$ is the present-day matter density parameter. We note that the spatial resolution of the $1024^3$ Cartesian grids is approximately $\Delta r_\text{grid} = 93\,\text{ckpc}$, corresponding to an equivalent spectral resolution of $\Delta v_\text{grid} = 9.3\,\text{km/s} \sqrt{(1+z)/7}$. This is sufficient for resolved IGM radiative transfer, as it is comparable to the thermal velocity in ionized regions, $v_\text{th} \approx 12.85\,T_4^{1/2}\,\text{km\,s}^{-1}$.

The Ly$\alpha$ absorption and emission characteristics are tracked through the neutral hydrogen absorption coefficient $k_\alpha$ and the conserved Ly$\alpha$ luminosity $L_\alpha$, respectively. The Cartesian grid includes three luminosity fields corresponding to different emission mechanisms: radiative recombination ($L_\alpha^\text{rec}$), collisional excitation ($L_\alpha^\text{col}$), and unresolved \HII regions around stars ($L_\alpha^\text{stars}$). In this paper, we focus on radiative transfer on IGM scales, so we choose to exclude emission from cells exceeding the equation of state density threshold. We also adopt an additional escape fraction for Ly$\alpha$ photons from stars of $f_\text{esc}^\alpha = 0.5$ to account for dust absorption prior to escaping into the IGM. While this value depends on complex radiative transfer phenomena such as the dust content, bulk velocity, and geometry for escape, our fiducial value is a conservative estimate for most lower-mass galaxies \citep[e.g.][]{Smith2019,Laursen2019,Garel2021,Bhagwat2024}.

The frequency-integrated absorption coefficient for each cell is $k_\alpha = n_{\HI} \,\sigma_\alpha$, where $n_{\HI}$ is the neutral hydrogen number density, and the bolometric cross section is $\sigma_\alpha = f_{12} \pi e^2 / (m_e c)$. A typical ray propagates through the medium experiencing an effective traversal opacity based on the volume-averaged absorption coefficient within each spectral grid cell:
\begin{equation} \label{eq:k_alpha}
  \langle k_\alpha \rangle = V_\text{grid}^{-1} \sum_i V_i k_{\alpha,i} = f_{12} \frac{\pi e^2}{m_e m_\text{H} c} \langle \rho_{\text{H\,\textsc{i}}} \rangle \, ,
\end{equation}
where $V_\text{grid}$ is the volume of the Cartesian grid cell, $V_i$ is the volume of each Voronoi cell, and the final equality relates Ly$\alpha$ absorption to the neutral hydrogen density $\rho_{\HI}$ from the \thesan outputs.

The specific details of the treatment of the emission mechanisms are described in \citet{Smith2022}, but here we briefly outline how each luminosity component is obtained. The recombination luminosity $L_\alpha^\text{rec}$, which results from the recombination cascade of ionized hydrogen, is given by:
\begin{equation}
    L_\alpha^\text{rec} = h\nu_0 \int P_\text{B}(T)\alpha_\text{B}(T) n_e n_p\,\text{d}V \, ,
\end{equation}
where $h$ is Planck's constant, $P_\text{B}(T) \approx 0.68$ is the probability of Ly$\alpha$ conversion per recombination event under Case B conditions (optically thick clouds), $\alpha_\text{B}$ is the Case B recombination coefficient, and $n_{e}$ and $n_{p}$ are the free electron and proton number densities, respectively. The collisional excitation luminosity $L_\alpha^\text{col}$, arising from the de-excitation of collisionally-excited hydrogen, is calculated as:
\begin{equation}
    L_\alpha^\text{col} = h\nu_0 \int q_{\rm{1s2p}}(T) n_{e}n_{\HI}\,\text{d}V \, ,
\end{equation}
where $q_{\rm{1s2p}}(T)$ is the rate coefficient for collisional excitation from the ground state to the first excited state. Finally, the nebular luminosity $L_\alpha^\text{stars}$, from unresolved \HII regions around stars, is determined by:
\begin{equation}
    L_\alpha^\text{stars} = 0.68 h\nu_0 (1 - f^{\rm{ion}}_{\rm{esc}}) \dot{N}_{\rm{ion}} \, ,
\end{equation}
where $f^{\rm{ion}}_{\rm{esc}}$ is the escape fraction of ionizing photons (calibrated to match constraints on the global reionization history), and $\dot{N}_{\rm{ion}}$ is the emission rate of ionizing radiation from stars. The details of obtaining the emissivity coefficients are discussed in Section~\ref{subsec::hubbleflow}.

\begin{figure*}
    \centering
    \includegraphics[width=0.9\textwidth]{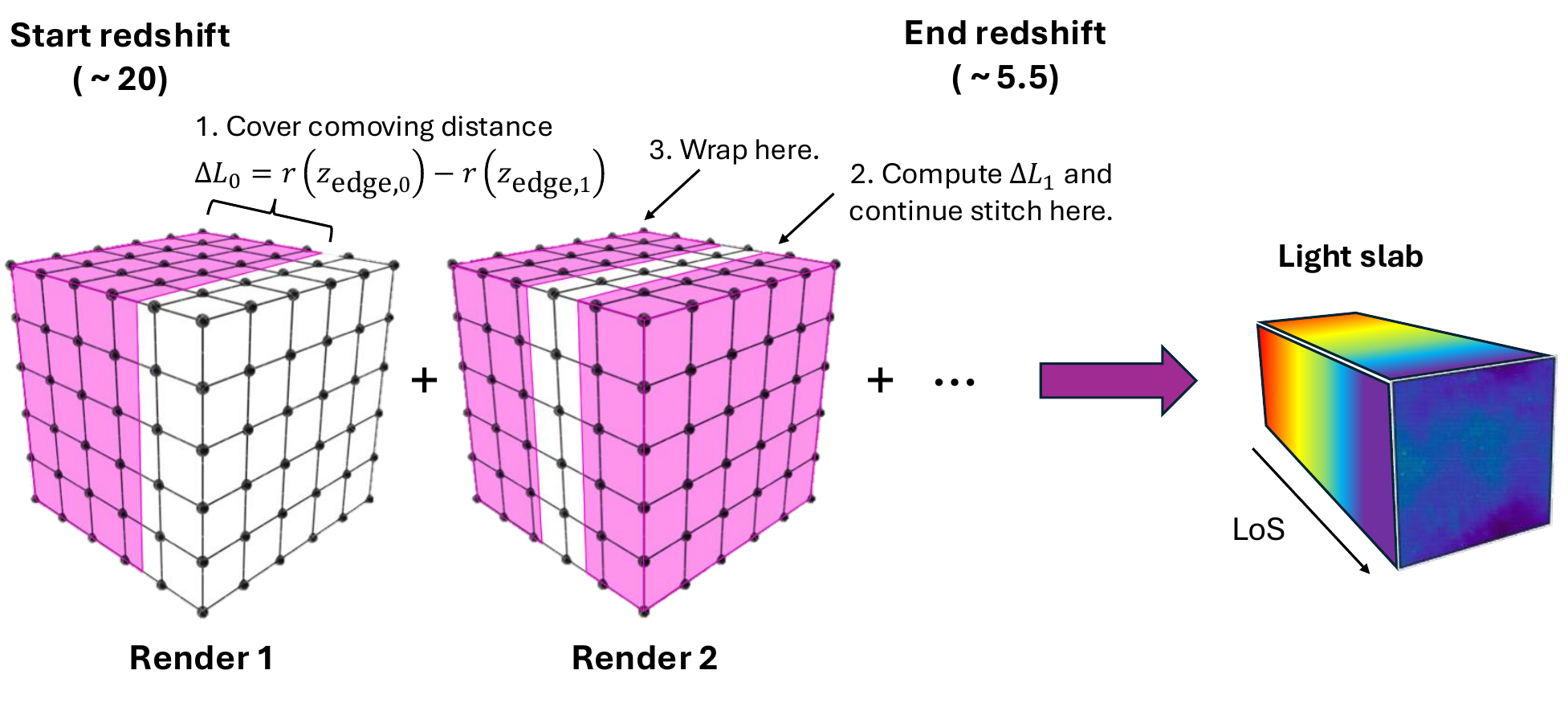}
    \caption{A schematic illustrating the process of stitching renders together to form a continuous light slab. \textbf{Step 1}: The leftmost box shows the first render as a 3D cube, and the shaded pink area corresponds to the data included in our stitched product, sliced along the third (LoS) dimension according to the allocated comoving distance range $\Delta L_0$. \textbf{Step 2}: For the second render, we continue from the point we had stopped at in the first render to ensure continuity along the LoS axis. \textbf{Step 3}: If $\Delta L_1$ exceeds the remaining length in the third dimension, we wrap around the box in the second render to cover the appropriate redshift/comoving distance range. This process is repeated for all subsequent cartesian renders until the desired redshift range is fully covered.}
    \label{fig:lightcuboid}
\end{figure*}

\subsection{Ly\texorpdfstring{$\balpha$}{α} radiative transfer}
\label{subsec::RT}
We are interested in obtaining the spatial fluctuations in the integrated Ly$\alpha$ specific intensity across a square patch of sky. To achieve this, we employ the frequency-dependent radiative transfer (RT) equation along the line of sight (LoS):
\begin{equation} \label{eq:OG_RT}
  \frac{\text{d}I_\nu}{\text{d}\ell} = j_\nu - k_\nu I_\nu
  \qquad \text{or} \qquad
  \frac{\text{d}I_\nu}{\text{d}\tau_\nu} = S_\nu - I_\nu \, ,
\end{equation}
where $I_\nu$ is the specific intensity at frequency $\nu$, $\ell$ is the path length along the LoS, $j_\nu$ is the emission coefficient, and $k_\nu$ is the absorption coefficient. As is shown in the second equation, we can alternatively express the RT equation in terms of the optical depth via $\text{d}\tau_\nu = k_\nu\text{d}\ell$, and where $S_\nu \equiv j_\nu / k_\nu$ is the source function. In this analysis, we neglect scattering back into the LoS and focus solely on absorption and emission processes, leaving investigations that include scattering for future work \citep[for a recent study see][]{AmVi2025}. The RT equation thus depends on the absorption and emission properties along the LoS, characterized by $k_\nu$ and $j_\nu$, respectively.

The formal solution to the RT equation over a path length $\ell$ within a given cell of spatially-uniform conditions with constant $S_\nu$ is:
\begin{equation} \label{eq:RT}
  I_\nu(\tau_\nu) = I_{\nu,0}\,e^{-\tau_\nu} + S_\nu \left(1 - e^{-\tau_\nu} \right) \, ,
\end{equation}
where $I_0$ is the initial specific intensity at the beginning of the path segment, and $\tau = k_\nu \ell$ is the optical depth. The first term represents the attenuation of the incoming radiation due to absorption, while the second term accounts for the emission along the path. However, in our cosmological integrations--discussed further in Section~\ref{subsec::hubbleflow}--we adopt absorption and emission profiles with a spatially-dependent term to account for Doppler shifting due to Hubble flow; we are hence unable to use Equation~\ref{eq:RT} directly and must solve the radiative transfer equation independently.

In any case, by iteratively applying our solution at each step $\ell$ along the LoS and assuming a specific Ly$\alpha$ line profile, we can compute the cumulative Ly$\alpha$ intensity $I(\nu)$. The spectral intensity can then be mapped to a redshift-dependent intensity $I(z)$ through the one-to-one relation $z = \frac{\nu_0}{\nu} - 1$ where $\nu_0 = 2.466 \times 10^{15}$\,Hz is the rest-frame frequency of the Ly$\alpha$ line, and $\nu$ is the observed frequency.

To most accurately model the Ly$\alpha$ line, we choose to adopt the Voigt profile, which captures both the Doppler (Gaussian) and the Lorentzian wings of the line, allowing for precise incorporation of spectral bins farther from the line center at a given redshift. We define the dimensionless frequency $x \equiv (\nu - \nu_0) / \Delta \nu_\text{D}$, where $\Delta \nu_\text{D} \equiv (v_\text{th}/c)\nu_0$ is the Doppler width of the profile and $v_\text{th} \equiv \sqrt{2 k_\text{B} T / m_\text{H}}$ is the thermal velocity. The Voigt profile is expressed using the dimensionless Hjerting--Voigt function, $H(x) = \sqrt{\pi} \Delta \nu_\text{D} \phi_\text{Voigt}(\nu)$:
\begin{equation} \label{eq:H}
  H(x) \equiv \frac{a}{\pi} \int_{-\infty}^\infty \frac{e^{-y^2}\text{d}y}{a^2+(y-x)^2} \, ,
\end{equation}
where $a \equiv \Delta \nu_L /2 \Delta \nu_D \approx 4.7 \times 10^{-4}\,T_4^{-1/2}$ is the so-called ``damping parameter'', and describes broadening relative to the natural line width $\Delta \nu_\text{L} = 9.936 \times 10^7$\,Hz, with $T_4 \equiv T / (10^4\,\text{K})$. Note that this damping parameter $a$ is different from the scale factor defined in Section ~\ref{subsec:renderfields}.

We perform a second-order expansion in $a$ on Eq.~(\ref{eq:H}), obtaining
\begin{equation}
\label{eq:secondorder}
    H(x) \approx e^{-x^2} \left( 1 + a^2 (1 - 2x^2) \right) + \frac{2 a}{\sqrt{\pi}} \left( 2 x F(x) - 1 \right) \, .
\end{equation}
where $F(x) \equiv \int_0^x e^{y^2 - x^2}\text{d}y$ is the Dawson integral. We further define the integral of this function as
\begin{equation} \label{eq:Upsilon}
    \Upsilon(x) \equiv \int H(x)\,\text{d}x \approx \frac{\sqrt{\pi}}{2} \text{erf}(x) - \frac{2 a}{\sqrt{\pi}} F(x) + a^2 x e^{-x^2} \, ,
\end{equation}
and the second integration of $H(x)$ over frequency as
\begin{align} \label{eq:int_upsilon}
    \bar{\Upsilon}(x) &\equiv \int \Upsilon(x) \, \text{d}x \\
    \approx &\frac{\sqrt{\pi}}{2} x\,\text{erf}(x)\,-\, \frac{a x^2}{\sqrt{\pi}} {}_2F_2\left(1 ,1 \, ; \, \frac{3}{2} , 2 \, ; \, -x^2\right) + \frac{(1 - a^2)}{2}e^{-x^2} \, , \notag
\end{align}
where ${}_2F_2$ is a generalized hypergeometric function.

\subsection{Performing LoS integrations with \thesan}
\label{subsec:stitching}
To perform LoS integrations using data from cosmological simulations \citep{Chan2019}, we construct a continuous data product spanning from the starting to ending redshifts of our simulation. This involves remapping the Cartesian data to a light-cone representation that we can ultimately integrate through to obtain different observables. We produce the desired light-cone by stitching the Cartesian grids together. However, in our approach, we produce a ``light slab'' instead of a true light-cone, assuming that the rays run parallel through our simulation, resulting in a flat 2D output.

In order to stitch together the Cartesian grids saved throughout the \thesan run, we use the configured render redshifts $[z_{0},z_{1},z_{2},..., z_{\rm{R}}]$ as midpoints. From these, we determine the redshift ``edges'' at which we transition from stitching one render to the next. Then, we compute the comoving distance corresponding to the allocated redshift range for each snapshot. For example, for the pair of edges assigned to the first render [$z_{\rm{edge,0}},z_{\rm{edge,1}}$] (as illustrated in the leftmost cube in Figure~\ref{fig:lightcuboid}), the comoving distance is:
\begin{equation}
  \Delta L_{0} = \int_{z_{\rm{edge,0}}}^{z_{\rm{edge,1}}} \frac{c\,\text{d}z}{H(z)} \, .
\end{equation}
We then slice the data within each render along the third spatial dimension (corresponding to the LoS) according to the computed comoving distance. This allows us to convert the distance to each pixel edge along the LoS into $\Delta z$ segments for our integration. To ensure continuity, when moving to the second snapshot, we start at the same index in the third spatial dimension where we stopped in the first snapshot. We repeat the process of computing the comoving distance covered between the redshift edges [$z_{\rm{edge},\,1}$, $z_{\rm{edge},\,2}$]. If the comoving distance range $\Delta L$ is greater than the remaining length in the box, as shown in the second cube in Figure~\ref{fig:lightcuboid}, we wrap around the box and continue from there. The \thesan box is periodic on all faces, so wrapping around in this manner is appropriate. We continue this process until we reach the last snapshot, and end up with a final data product that runs from the starting to the ending redshifts for each pixel $(p,q) \in N\times N$, allowing us to integrate continuously along the redshift dimension. 

The continuous wrapping in this procedure introduces a periodicity that begs the questions: how long does light travel between each rendering? And how many times do we cross the box in constructing our light slab? Assuming a matter-dominated Universe, we have:
\begin{align}
  &\Delta x_{n+1} \equiv x(a_{n+1}) - x \approx \frac{2 c}{H_0 \sqrt{\Omega_0}} \left[ \sqrt{a_2} - \sqrt{a_1} \right] \approx \frac{c (a_{n+1} - a)}{H_0 \sqrt{\Omega_0} \sqrt{a}} \, . \notag \\
  &\quad \Rightarrow \quad \frac{\Delta x}{100\,\text{cMpc}} \approx 19.9\,\left( \sqrt{\frac{6.5}{1+z_2}} - \sqrt{\frac{14}{1+z_1}} \right) .
\end{align}
So at $z=6$, for example, light travels $\sim 4$\, cMpc between renderings. From the second equation above, we cross the box 20 times going from redshift $13$ to $5.5$. While aliasing the same structures multiple times is not ideal, the setup serves as a good benchmark and approximates the first-order effects well \citep[e.g.][]{Konietzka2025}.

We will use the product of the above procedure to generate the intensity maps and power spectra that are essential to predict and interpret observations from upcoming Ly$\alpha$ LIM experiments.

\subsection{Cosmological radiative transfer integration}
\label{subsec::hubbleflow}

In this Section, we detail the possible approaches to solving the RT equation when including only absorption sources, only emission sources, and both emission and absorption together. 

The possibilities for each of the base assumptions for our integration scheme are provided below, showing how we adopt a series of improvements to accurately capture radiative transfer effects. In Table~\ref{tab:integration_schemes}, we provide the mathematical representations of the various integration schemes presented in the rest of this Section; throughout this paper, we also explore the effects of incorporating these improvements on the results. 
\begin{itemize}
    \item \textbf{Voigt Profile:} To allow for thermal and turbulent broadening of the line and damping wing absorption, we choose to adopt the full Voigt profile rather than a delta-function representation.
    \item \textbf{Band Averages:} To conserve the number of emitted photons, it is ideal to integrate over frequency bands instead of adopting a discrete sampling of frequencies.
    \item \textbf{Comoving Frame:} It is advantageous to employ continuous Doppler shifting to avoid line skipping in the absorption treatment rather than piecewise constant static integrations.
    \item \textbf{Emission Spectra:} To explore the impact of unresolved sub-galactic reprocessing, we would like to allow for redshifted emission, or other more general spectral profiles for the emissivity.
\end{itemize}
Unfortunately, it is not possible to include all of these effects when looking for an analytically tractable solution, so, depending on scientific priorities and numerical sensitivity, we suggest appropriate combinations of these treatments to harness their advantages or avoid shortcomings.

\subsubsection{Absorption}

To perform cosmological radiative transfer, we must account for the effects of Doppler shifting due to the continuous Hubble flow. A change in velocity $\Delta v$ due to Hubble flow expansion induces a wavelength shift given by $\Delta\lambda = \lambda_0 \Delta v / c$,
or equivalently, for a path length $\ell$, using the relation between the dimensionless Doppler frequency and the recessional velocity, $x = -\Delta v / v_\text{th}$, we have:
\begin{equation}
  x' = x - \mathcal{K} \ell \quad \text{with} \quad \mathcal{K} \equiv \frac{H(z)}{v_\text{th}} \, .
\end{equation}
Following the treatment in \citet{Smith2022,Smith2025}, with a second order expansion in $a$ (as shown in Eq.~\ref{eq:secondorder}) the traversed optical depth over a proper segment length $\ell$ is given by solving Eq.~(\ref{eq:OG_RT}) with explicit Doppler shifting $\text{d}I_\nu / \text{d}\ell = -k_0\,H(x - \mathcal{K} \ell) I_\nu$:
\begin{equation} \label{eq:abs-voigt-comb}
  \tau_\nu \equiv k_0 \int_0^\ell H\left( x - \mathcal{K} \ell' 
  \right)\,\text{d}\ell' \approx \frac{k_0}{\mathcal{K}} \big[\Upsilon(x) - \Upsilon(x-\mathcal{K}\ell)\big] \, ,
\end{equation}
where $k_0 \equiv n_\text{\HI} \sigma_0$ is the absorption coefficient at line center, with a cross-section $\sigma_0 = f_{12} \sqrt{\pi} e^2 / (m_e c \Delta \nu_\text{D})$ and oscillator strength $f_{12} = 0.4162$. We can equate this cross-section to that in Eq.~(\ref{eq:k_alpha}) to relate the absorption coefficient at line center $k_0$ to the frequency-integrated absorption coefficient $k_\alpha$ obtained from the renders:
\begin{equation}
  k_{0} = \left( \frac{c}{\sqrt{\pi}\nu_{\rm{th}} \nu_0}\right) k_\alpha \, .
\end{equation}
As expected, $k_0$ and $k_\alpha$ differ by a factor with units of frequency.
We introduce the following notation for frequency band averaging:
\begin{equation}
\label{eq:band}
  \langle f \rangle \equiv \frac{1}{\Delta x_i} \int_{x_{i}}^{x_{i+1}} f(x)\,\text{d}x \, .
\end{equation}
While Eq.~(\ref{eq:abs-voigt-comb}) gives us the optical depth for ``comb samplings'' in frequency, but we can also obtain the band-averaged equivalent over some interval $[x_i, x_{i+1}]$ where $\Delta x_i \equiv x_{i+1} - x_{i} > 0$:
\begin{equation}
\label{eq:abs-voigt-band}
  \langle \tau_\nu \rangle
  = \frac{k_0}{\mathcal{K}\,\Delta x_i} \bar{\Upsilon}(x') \Big|^{x}_{x - \mathcal{K} \ell } \Big|^{x_{i+1}}_{x_i} \, .
\end{equation}
Integrating along the LoS over a proper length $\ell$, we obtain the resulting intensity from $\text{d}I_\nu / \text{d}\ell = -k_0\,H(x - \mathcal{K} \ell) I_\nu$:
\begin{equation}
\label{eq:Inu-abs-voigt-band}
  I\,_\nu^{\rm{abs}} = I_{\nu,0}\,e^{-\tau_\nu} \quad \text{and} \quad \langle I\,_\nu^{\rm{abs}}\rangle = \left\langle I_{\nu,0}\,e^{-\tau_\nu} \right\rangle \geq \langle I_{\nu,0}\rangle\,e^{-\langle\tau\rangle_\nu} \, ,
\end{equation}
where the last expression adopts a prolongation approximation decoupling the unresolved product within a coarse spectral bin and applies Jensen's Inequality for convex functions. In practice, we use the final version as an analytically tractable lower limit, treating sub-band structure in an average sense. Similarly, if we adopt the Dirac delta function for our absorption, $k_\nu = \sqrt\pi k_0\,\delta(x - \mathcal{K}\ell)$, we straightforwardly compute the optical depth as:
\begin{equation}
\label{eq:abs-delta-comb}
    \tau_\delta \equiv \sqrt{\pi}\,k_0 \int_0^{\ell} \delta(x-\mathcal{K}\ell')\,d\ell' =
    \frac{\sqrt{\pi}\,k_0}{\mathcal{K}} \, \mathds{1}_{[0,\ell]}(\ell_\text{abs}) \, ,
\end{equation}
where we define $\ell_\text{abs} \equiv x / \mathcal{K}$ and the indicator function is defined as
\begin{equation}
  \mathds{1}_A(x) \equiv \begin{cases} 1 & x \in A \\ 0 & \text{otherwise} \end{cases} \, .
\end{equation}
By introducing the convenient shorthand clipping notation for the fraction of the band falling within the delta function conditions
\begin{equation}
 \mathcal{F}_{\;a}^{b} \equiv \Delta x_i^{-1} \left[ \min(\max(x_{i+1}, a), b) - \min(\max(x_i, a), b) \right] \, ,
\end{equation}
the band-averaged equivalent optical depth of $\langle \tau \rangle_\delta = \frac{\sqrt{\pi}\,k_0}{\mathcal{K}} \, \mathcal{F}_{\;0}^{\mathcal{K}\ell}$. However, in this case, the intensity has an exact analytic solution:
\begin{align}
\label{eq:intensity-abs-delta-band}
    \langle I\,_\delta^{\rm{abs}} \rangle &\equiv \left\langle I_{\delta,0}\,e^{-\tau_\delta}\right\rangle \approx \langle I_{\delta,0}\rangle\,\left\langle e^{-\tau_\delta}\right\rangle \notag \\
    &= \langle I_{\delta,0} \rangle\,\left(1 + \mathcal{F}_{\;0}^{\mathcal{K}\ell}\,\left[\exp\left(-\frac{\sqrt{\pi}\,k_0}{\mathcal{K}}\right) - 1 \right] \right) \, ,
\end{align}
where we again average over the sub-band structure of $\langle I \rangle_{\delta,0}$.

For some of our analyses, we are interested in evaluating purely the absorption characteristics of the IGM along the LoS. Since our data are projected onto a 3D Cartesian grid, integrating across the LoS dimension leads to a spectral transmission computed for each pixel in our projected 2D image. Given a 3D grid resolution $N^{3}$, we denote our frequency/redshift-dependent transmission as $\mathcal{T}_{N\times N}(\nu) \rightarrow \mathcal{T}_{N\times N}(z_{\rm{Ly}\alpha})$, using the straightforward mapping from frequency to redshift space. Hence, for a given pixel $(i$,\,$j)$ in our image, we can obtain the frequency-dependent transmission from the cumulative optical depth $\tau(\nu)$ or band-averaged intensity $\langle I \rangle$:
\begin{equation}
  \mathcal{T}(\nu) = \exp\left[ -\sum_{m=1}^M \tau_{m}(\nu) \right] \quad \text{and} \quad \langle\mathcal{T}\rangle(\nu) = \prod_{m=1}^M \frac{\langle I \rangle_m}{\langle I \rangle_{0,m}} \, ,
\end{equation}
where $\tau_{m}(\nu)$ is the optical depth contributed by the $m^\text{th}$ segment along the LoS, computed using Eqs.~(\ref{eq:abs-voigt-comb}) and~(\ref{eq:abs-delta-comb}), band-averaged intensities are taken from Eqs.~(\ref{eq:Inu-abs-voigt-band}) and (\ref{eq:intensity-abs-delta-band}), and $M$ is the total number of segments traversed. Note that $M$ here is not simply the resolution $N$, because we stitch together all of the renders within our simulation to form a coherent LoS from the start to the end redshift; hence, $M \gg N$. Since the on-the-fly simulation adopted a uniform Cartesian grid for the renderings, $M$ turns out to be the same for all of the pixels in our final image. The details of this stitching process are discussed in Section~\ref{subsec:stitching}.

\subsubsection{Emission}

We are also interested in the frequency-dependent emissivity for each of our emission sources $j_\nu \equiv \text{d}E / (\text{d}t\text{d}V\text{d}\nu\text{d}\Omega)$, given by the integral $\mathcal{L} = \iiint j_\nu \text{d}V\text{d}\nu\text{d}\Omega$, recovering the luminosity in units of $\text{erg\,s}^{-1}$. When including continuous Hubble flow and assuming a Voigt profile, we have in units of $\text{erg\,s}^{-1}\,\text{cm}^{-2}\,\text{sr}^{-1}\,\text{Hz}^{-1}$:
\begin{equation}
  j_\nu = j_0 H(x - x_\text{out} - \mathcal{K} \ell) \quad \text{where} \quad j_0 \equiv \frac{\mathcal{L}}{4\pi\Delta A\Delta\ell\sqrt{\pi}\,\Delta\nu_D} \, ,
\end{equation}
with $\Delta A$ the physical pixel size, and the spatial, frequency, and solid angle components of the emissivity included in the last equality are given by $1/(\Delta A \Delta \ell)$, $H(x) / (\sqrt{\pi} \Delta \nu_\text{D})$, and $1/(4\pi)$, respectively. For photons from unresolved \HII regions, we incorporate an additional Doppler shift $x_\text{out}$, which is the unitless analog for the Doppler shift induced by sub-resolution outflows in the form of $\Delta\lambda/\lambda_0 = \Delta v_{\rm{out}}/c$, where $\Delta v_{\rm{out}}$ is the assumed expansion velocity of the outflow \citep{VeSch2006}. Using the definition of the unitless frequency, we have $x_\text{out} = -\frac{\Delta\lambda}{\lambda_{0}}\frac{c}{v_{\rm{th}}} = -\Delta v_{\rm{out}}/ v_{\rm{th}}$. For photons originating from recombination or collisional excitation, and when \HII regions are assumed to have no initial redward boost from expanding outflows, we simply set $x_\text{out}=0$. When averaged over some frequency band $x \in [x_{i},x_{i+1}]$ we get
\begin{equation}
  \langle j_\nu \rangle
  = j_0 \frac{\Upsilon(x-x_\text{out}-\mathcal{K}\ell)}{\mathcal{K}\Delta x_i} \bigg|^{x_{i+1}}_{x_{i}} \, .
\end{equation}
Integrating along the LoS over a proper length $\ell$, we obtain the resulting intensity from $\text{d}I_\nu^{\rm{em}} / \text{d}\ell = j_0 H(x - x_\text{out} - \mathcal{K} \ell)$:
\begin{equation}
\label{eq:em-voigt-comb}
  I_\nu^{\rm{em}} = I_{\nu,0} + \frac{j_0}{\mathcal{K}} \Upsilon(x') \Big|^{x-x_{\text{out}}}_{x - x_\text{out} - \mathcal{K} \ell} \, .
\end{equation}
Such a solution is appropriate for the case that all Ly$\alpha$ photons are scattered back into the LoS.
Similarly, when integrated over the path we can compute the band-averaged spectral intensity as
\begin{equation}
\label{eq:em-voigt-band}
  \langle I\,_\nu^{\rm{em}} \rangle = \langle I_{\nu,0} \rangle + \frac{j_0}{\mathcal{K}\,\Delta x_i} \bar{\Upsilon}(x') \Big|^{x-x_{\text{out}}}_{x - x_\text{out} - \mathcal{K} \ell} \,\Big|^{x_{i+1}}_{x_i} \, .
\end{equation}
Finally, we derive the equivalent forms assuming $j_\nu = \sqrt{\pi}\,j_0\,\delta(x - x_\text{out} - \mathcal{K} \ell)$, i.e., a unit impulse at line center. For comb sampling in frequency, the path integrated emission-only intensity becomes
\begin{equation}
\label{eq:em-delta-comb}
    I\,_\delta^{\rm{em}} = I_{\delta,0} + \frac{\sqrt{\pi}\,j_0}{\mathcal{K}} \, \mathds{1}_{[0,\ell]}(\ell_\text{src}) \, ,
\end{equation}
where $\ell_\text{src} \equiv (x - x_\text{out}) / \mathcal{K}$,
and integrated over [$x_{i},x_{i+1}$], we have:
\begin{equation}
\label{eq:em-delta-band}
    \langle I\,_\delta^{\rm{em}} \rangle = \langle I_{\delta,0} \rangle + \frac{\sqrt{\pi}\,j_0}{\mathcal{K}}\,\mathcal{F}_{\;x_\text{out}}^{x_\text{out} + \mathcal{K}\ell} \, .
\end{equation}

\begin{table*}
  \centering
  \caption{Each emission/absorption profile used in this work, with $j_\nu$ and $k_\nu$ specified as either Dirac delta $\delta(x)$ or Voigt profile $\varphi(x) = H(x)/\sqrt{\pi}$. The last two columns list the corresponding comb-sampled and band-averaged specific intensities, together with references to the defining equations. All radiative transfer (RT) solutions include cosmological redshifting ($\mathcal{K}\ell$).}
  \label{tab:integration_schemes}
  \addtolength{\tabcolsep}{-2pt}
  \renewcommand{\arraystretch}{1.12}

\begin{tabular}{p{0.15\textwidth} p{0.15\textwidth} p{0.15\textwidth} p{0.2\textwidth} p{0.2\textwidth}}
    \hline
    Classification & Emission ($j_\nu$) & Absorption ($k_\nu$) & Comb & Band \\
    \hline
    \multirow{2}{*}{Absorption only}
      & 0   & $\delta(x)$   & $I_{\delta}^{\rm abs}$ \; (Eq.~\ref{eq:abs-delta-comb}) 
 & $\langle I_{\delta}^{\rm abs} \rangle$ \; (Eq.~\ref{eq:intensity-abs-delta-band}) \\[2pt]
      & 0     & $\varphi(x)$   & $I_{\nu}^{\rm abs}$ \; (Eq.~\ref{eq:abs-voigt-comb}) & $\langle I_{\nu}^{\rm abs} \rangle$ \; (Eqs.~\ref{eq:abs-voigt-band}, \ref{eq:Inu-abs-voigt-band}) \\[4pt]

    \hline
    \multirow{2}{*}{Emission only}
      & $\delta(x - x_{\rm out})$ & 0                       & $I_{\delta}^{\rm em}$ \; (Eq.~\ref{eq:em-delta-comb}) & $\langle I_{\delta}^{\rm em} \rangle$ \; (Eq.~\ref{eq:em-delta-band}) \\[2pt]
      & $\varphi(x - x_{\rm out})$ & 0                     & $I_{\nu}^{\rm em}$ \; (Eq.~\ref{eq:em-voigt-comb}) & $\langle I_{\nu}^{\rm em} \rangle$ \; (Eq.~\ref{eq:em-voigt-band}) \\[4pt]
    \hline
    \multirow{3}{*}{Absorption + Emission}
      & $\delta(x - x_{\rm out})$ & $\delta(x)$             & $I_{\delta}$ \; (Eq.~\ref{eq:delta-comb-absem})  & $\langle I_{\delta} \rangle$ \; (Eq.~\ref{eq:delta-band-absem}) \\[2pt]
 & $\delta(x - x_{\rm out})$ & $\varphi(x)$ & $I_{\delta\varphi}$ \; (Eq.~\ref{eq:dv-comb-absem})  & $\langle I_{\delta\varphi} \rangle$ \; (Eq.~\ref{eq:dv-band-absem}) \\[2pt] & $\varphi(x)$ & $\varphi(x)$           & $I_{\nu}$ \; (Eq.~\ref{eq:absem_voigt_comb})  & $\langle I_{\nu} \rangle$ \; (Eq.~\ref{eq:absem_voigt_band}) \\[4pt]
    \hline 
    No self-absorption
      & $\varphi(x - x_{\rm out})$ & $\varphi(x)$        & $I_{\nu}^{\rm out}$ \; (Eq.~\ref{eq:voigt-noSA})  & $\langle I_{\nu}^{\rm out} \rangle$ \; (Eq.~\ref{eq:voigt-band-noSA}) \\[2pt]
    \hline
    \multicolumn{5}{l}{\parbox{.975\textwidth}{\textbf{Comments:} For Ly$\alpha$ transmission most methods perform well: $I_\delta$ is traditional and $I_\nu$ improves DLA features; in terms of band-integration, $\langle I_\delta \rangle$ is exact and allows lower resolution and $\langle I_\nu \rangle$ is an accurate lower limit. We do not include the $\delta(x-x_{\rm out})-\varphi(x)$ combination in the results as it turns out to perform similarly to the $\delta(x-x_{\rm out})-\delta(x)$ combination. Note that the $\varphi(x)-\varphi(x)$ absorption + emission combination does not have an analytic solution for the incorporation of a velocity offset, and our fiducial model for outflows is the band-integrated ``no self-absorption'' model $\langle I_{\nu}^{\rm out} \rangle$ (refer to Appendix~\ref{appendix:outflows}).}
    }
  \end{tabular}

  \addtolength{\tabcolsep}{2pt}
  \renewcommand{\arraystretch}{1.0}
\end{table*}

\subsubsection{Absorption and emission}
If we adopt the same functional forms for both the emission and absorption profiles and do not include a boost from outflows in the emission component (i.e., set $x_\text{out}=0$), we end up with the following equation when we have a Voigt Profile for both emission and absorption: $\text{d}I_\nu / \text{d}\ell = j_0 H(x - \mathcal{K} \ell) - k_0 H(x - \mathcal{K} \ell) I_\nu$, which transforms to $\text{d}I_\nu / \text{d}\tau_\nu = S_0 - I_\nu$ with $\tau_\nu$ given by Eq.~(\ref{eq:abs-voigt-comb}) and frequency-independent source function:
\begin{equation}
\label{eq:constant_source}
  S_0 \equiv \frac{j_0}{k_0} = \frac{\mathcal{L}}{4 \pi \Delta A\Delta\ell \Delta \nu_\text{D} k_0} \, , 
\end{equation}
and can directly use Eq.~(\ref{eq:RT}) to obtain the Ly$\alpha$ spectral intensity. More explicitly, when adopting a comb-sampled Voigt form, we have
\begin{equation}
\label{eq:absem_voigt_comb}
   I_\nu = I_{\nu,0}\,e^{-\tau_\nu} + S_0 \left(1 - e^{-\tau_\nu} \right) \, , 
\end{equation}
with $\tau_\nu$ defined in Eq.~(\ref{eq:abs-voigt-comb}). For the band-integrated Voigt, we have
\begin{equation}
\label{eq:absem_voigt_band}
  \langle I_\nu \rangle \geq \langle I_{\nu,0}\rangle\,e^{-\langle\tau_\nu\rangle} + S_0 \left(1 - e^{-\langle\tau_\nu\rangle}\right) \, , 
\end{equation}
It is also analytically tractable to replace one of these with a Dirac delta function. We find a practical and physically realistic choice to be sharp emission; with a boost from outflows, we must solve:
\begin{equation}
  \frac{\text{d}I_\nu}{\text{d}\ell} = \sqrt{\pi} j_0 \delta(x - x_\text{out} - \mathcal{K} \ell) - k(x - \mathcal{K} \ell) I_\nu \, ,
\end{equation}
which has a general solution of
\begin{equation}
  I = I_0\,e^{-\tau(\ell)} + \frac{\sqrt{\pi}\,j_0}{\mathcal{K}} \exp\left[ \tau(\ell_\text{src}) - \tau(\ell) \right] \, \mathds{1}_{[0,\ell]}(\ell_\text{src}) \, ,
\end{equation}
where the general optical depth is $\tau(\ell) \equiv \int_0^\ell k(x - \mathcal{K}\ell')\,\text{d}\ell'$. Specifically, adopting $\tau_\nu$ from Eq.~(\ref{eq:abs-voigt-comb}) leads to
\begin{align}
\label{eq:dv-comb-absem}
  &I_{\delta\varphi} = I_{\delta\varphi,0}\,\exp\left(\frac{k_0}{\mathcal{K}} \left[\Upsilon(x-\mathcal{K}\ell) - \Upsilon(x)\right]\right) \\
  &+ \frac{\sqrt{\pi}\,j_0}{\mathcal{K}} \exp\left( \frac{k_0}{\mathcal{K}}\left[ \Upsilon(x-\mathcal{K}\ell) - \Upsilon(x_\text{out}) \right]\right) \, \mathds{1}_{[0,\ell]}(\ell_\text{src}) \, . \notag
\end{align}
The lower-limit approximation for band-integration becomes
\begin{align}
\label{eq:dv-band-absem}
  &\langle I \rangle_{\delta\varphi} \geq \langle I\rangle_{\delta\varphi,0}\,e^{-\langle\tau\rangle_\nu} +  \frac{\sqrt{\pi}\,j_0}{\mathcal{K}} \exp\left( -\frac{k_0}{\mathcal{K}} \Upsilon(x_\text{out})\right) \\
  &\times \mathcal{F}_{x_\text{out}}^{x_\text{out}+\mathcal{K}\ell} \exp\left( \frac{k_0 \bar{\Upsilon}(x')\big|_{\max(x_i,x_\text{out}) - \mathcal{K}\ell}^{\min(x_{i+1},x_\text{out} + \mathcal{K}\ell) - \mathcal{K}\ell}}{\mathcal{K}\mathcal{F}_{x_\text{out}}^{x_\text{out}+\mathcal{K}\ell} \Delta x_i}\right) \, , \notag
\end{align}
noting that the emission term is only included if $\mathcal{F}_{x_\text{out}}^{x_\text{out}+\mathcal{K}\ell} > 0$. Similarly, adopting $\tau_\delta$ from Eq.~(\ref{eq:abs-delta-comb}), the exact comb solution is
\begin{equation}
\label{eq:delta-comb-absem}
  I_\delta = I_{\delta,0}\,e^{-\tau_\delta} + \frac{\sqrt{\pi}\,j_0}{\mathcal{K}} \, \mathds{1}_{[0,\ell]}(\ell_\text{src}) \, \exp\left( -\frac{\sqrt{\pi}\,k_0}{\mathcal{K}} \, \mathds{1}_\text{out} \right) \, ,
\end{equation}
where we define $\mathds{1}_\text{out} \equiv \mathds{1}_{[0,\infty)}(x_\text{out}) = \mathds{1}_{[0,\ell_\text{abs}]}(\ell_\text{src})$, i.e. the emission source occurs before the absorption. In our case, with outflows ($x_\text{out} < 0$) then $\mathds{1}_\text{out} = 0$. The exact band-integrated solution is:
\begin{align}
\label{eq:delta-band-absem}
  \langle I_\delta\rangle &= \langle I_{\delta,0} \rangle\,\left(1 + \mathcal{F}_{\;0}^{\mathcal{K}\ell}\,\left[\exp\left(-\frac{\sqrt{\pi}\,k_0}{\mathcal{K}}\right) - 1 \right] \right) \notag \\ &+ \frac{\sqrt{\pi}\,j_0}{\mathcal{K}}\,\mathcal{F}_{\;x_\text{out}}^{x_\text{out} + \mathcal{K}\ell} \, \exp\left( -\frac{\sqrt{\pi}\,k_0}{\mathcal{K}} \, \mathds{1}_\text{out} \right) \, . 
\end{align}
Finally, we also consider a model in which all Ly$\alpha$ photons are scattered back into the LoS locally but treat scattering as absorption outside the emitting segment. In this case, the Voigt emission and absorption comb-sampled spectral intensity is
\begin{equation}
\label{eq:voigt-noSA}
  I\,_\nu^{\rm{out}} = I_{\nu,0}\,e^{-\tau_\nu} + \frac{j_0}{\mathcal{K}} \Upsilon(x') \Big|^{x-x_{\text{out}}}_{x - x_\text{out} - \mathcal{K} \ell} \, ,
\end{equation}
and the band-averaged spectral intensity is
\begin{equation}
\label{eq:voigt-band-noSA}
  \langle I\,_\nu^{\rm{out}} \rangle = \langle I_{\nu,0}\rangle\,e^{-\langle\tau_\nu\rangle} + \frac{j_0}{\mathcal{K}\,\Delta x_i} \bar{\Upsilon}(x') \Big|^{x-x_{\text{out}}}_{x - x_\text{out} - \mathcal{K} \ell} \,\Big|^{x_{i+1}}_{x_i} \, ,
\end{equation}
where we denote the corresponding spectral intensity as $I_{\nu}^{\rm{out}}$ because this result ultimately serves as our best model for subgrid outflows (see Figure~\ref{fig:dv-combo}).
\begin{figure}
    \centering
    \includegraphics[width=\linewidth]{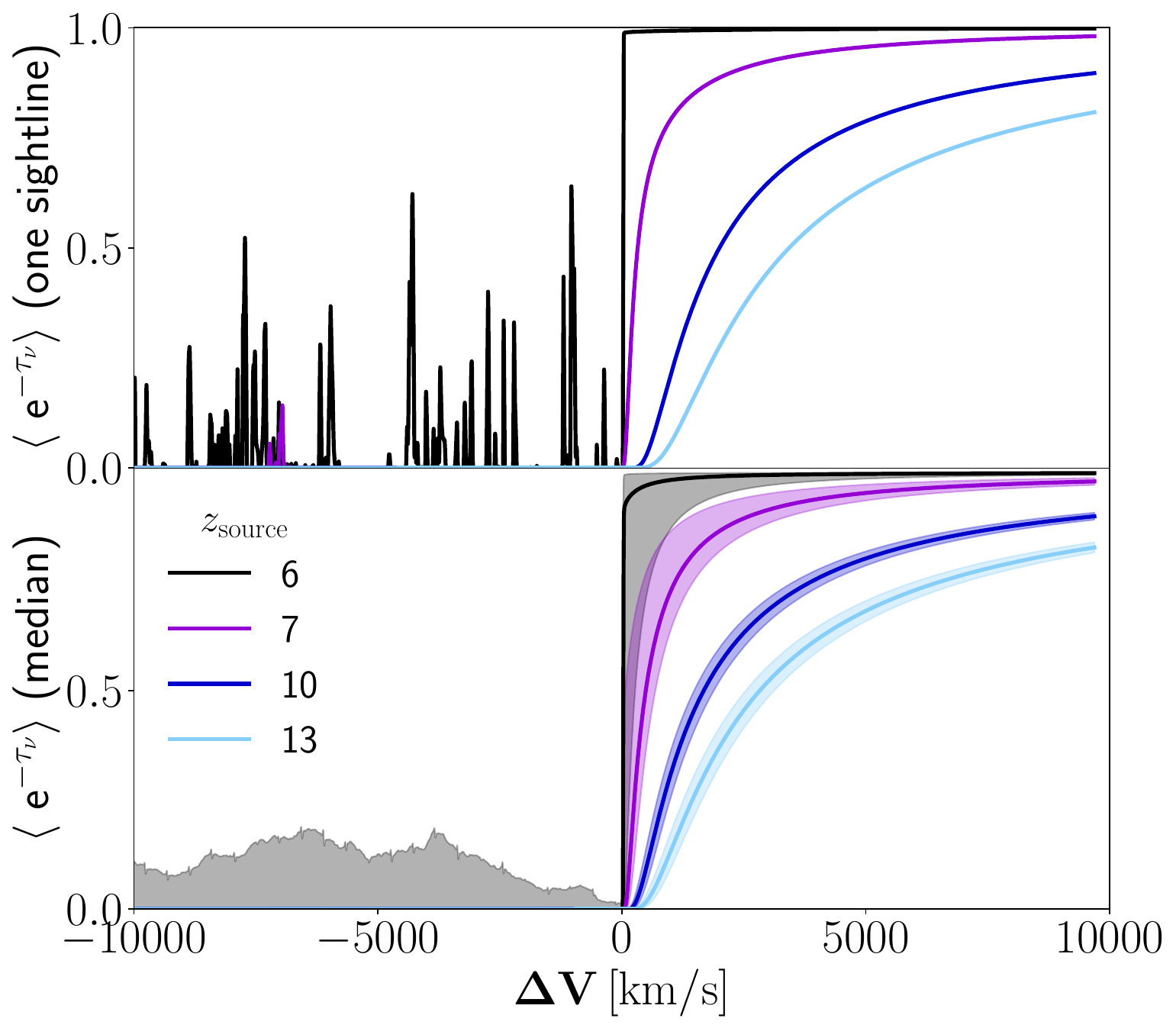}
    \caption{Ly$\alpha$ damping wing profiles, both along a random sightline (top panel) and when taking the median and $1\sigma$ deviation (bottom panel) of our entire $1024\times1024$ grid for different source redshifts: $z_{\rm{source}} \in \{6, 7, 10, 13\}$ in black, purple, dark blue, and light blue respectively. In the bottom panel, the $1\sigma$ deviation is plotted as a shaded region in the respective color. The strength of the absorption increases with redshift due to an increase in neutral hydrogen in the IGM. By $z_{\rm{source}}=6$, transmission peaks appear blueward of the line center and the characteristically broad red wing absorption is replaced with a sharp cutoff sensitive to the local environment around the source. Correspondingly, the increased scatter is a result of a more patchy reionization morphology during the later stages of reionization.}
    \label{fig:dampingwing_combined}
\end{figure}
We compute the spectral intensity for each pixel in our 2D image, $\mathcal{I}_{N\times N}(\nu)$. Since the RT equation requires both attenuating past radiation flux and adding flux contributions from the source function at that position, we must update $\mathcal{I}_{N\times N}(\nu)$ at each segment along the integration path. More specifically, if we take any of our solutions (save for the band-integrated $\delta$--$\delta$ solution, which has a slightly different form) we can update the spectral intensity at a pixel $(i,j)$ at some position along the LoS path indexed by $m$ as follows:
\begin{equation}
    \mathcal{I}_{m}(\nu) = \Theta_{m} + I_{m-1}(\nu) \, e^{-\tau'_m(\nu)} \, ,
\end{equation}
where $I_{m-1}$ is the total incoming intensity, i.e., the computed spectral intensity at that pixel for the previous segment and $\Theta_m$ denotes the contribution from the source function, which both depend on the absorption/emission profiles that we adopt. For Eq.~(\ref{eq:constant_source}), for example, which applies to the $\varphi$--$\varphi$ pairing, we have:
\begin{equation}
\label{eq:computation_constantSnu}
    \Theta_m =
        S_m\,\left(1 - e^{-\tau'_m(\nu)}\right) \, .
\end{equation}
Note that $\tau'_m(\nu) = \tau_m(x-\mathcal{K}\ell) - \tau_m(x)$. This procedure is repeated for each step along the LoS, and the reference frequency $x$ is continuously redshifted after each integration.

At cosmological distances, ``surface brightness dimming'' becomes important \citep{To1930,To1934}. The observed intensity will hence dampen the emitted value through two effects: first, the luminosity-to-angular diameter distance ratio squared leads to a $(1+z)^4$ reduction via $d_\text{L} = (1 + z)^2\,d_\text{A}$, and second, the spectral binning is modified via $\Delta \lambda_\text{obs} = (1+z)\,\Delta \lambda_\text{em}$ and $\Delta \nu_\text{obs} = \Delta \nu_\text{em} / (1+z)$. In this formulation, the surface brightness (SB) in units of $\text{erg\,s}^{-1}\,\text{cm}^{-2}\,\text{sr}^{-1}$ is:
\begin{equation}
    \text{SB} = \frac{\mathcal{L}}{4 \pi d_\text{L}^2 \Omega_\text{pix}} = \frac{\mathcal{L}}{4 \pi (1+z)^4 \Delta A} \, ,
\end{equation}
and the observed intensity in units of $\text{erg\,s}^{-1}\,\text{cm}^{-2}\,\text{sr}^{-1}\,\text{Hz}^{-1}$ is:
\begin{equation}
    I_{\nu,\text{obs}} = \frac{I_{\nu,\text{emit}}}{(1+z)^3} \, .
\end{equation}
We can thus account for this effect by simply dividing the source function by $(1+z)^3$ at each step.

\begin{table}
\caption{Average transmission for two transmission models, $\langle e^{-\tau_{\delta}}\rangle$ is $\langle I\,^{\rm{abs}}_{\delta} \rangle$ from Eq.~(\ref{eq:intensity-abs-delta-band}) and  $\langle e^{-\tau_{\nu}}\rangle$ is $\langle I\,_{\nu}^{\rm abs} \rangle$ from Eqs.~(\ref{eq:abs-voigt-band}) and (\ref{eq:Inu-abs-voigt-band}), at redshifts $5.6 \leq z \leq 6.2$ ($\Delta z = 0.1$). Values are shown in per cent. Each row corresponds to a model applied at that redshift. We adopt a smaller sample of sightlines to emulate \citet{BoDa22}, and a larger sample size to ensure convergence.}
\addtolength{\tabcolsep}{4pt}
\begin{tabular}{cccc}
\hline
Redshift & \quad $\langle e^{-\tau_\delta}\rangle$ [\%] \quad & \quad $\langle e^{-\tau_\nu}\rangle$ [\%] \quad & \# Sightlines \\
\hline
5.6 & $23.2 \pm 5.27$ & $12.1 \pm 5.55$ & 33 \\
    & $23.8 \pm 5.61$ & $13.1 \pm 4.85$ & 381 \\
5.7 & $21.4 \pm 4.40$ & $10.2 \pm 4.46$ & 49 \\
    & $21.3 \pm 4.49$ & $11.5 \pm 4.04$ & 384 \\
5.8 & $17.9 \pm 5.07$ & $7.01 \pm 4.81$ & 40 \\
    & $17.9 \pm 4.64$ & $8.48 \pm 4.57$ & 393 \\
5.9 & $18.2 \pm 3.90$ & $7.21 \pm 3.42$ & 49 \\
    & $17.6 \pm 3.63$ & $8.76 \pm 3.50$ & 428 \\
6.0 & $13.4 \pm 3.55$ & $3.18 \pm 3.15$ & 42 \\
    & $13.6 \pm 3.65$ & $4.66 \pm 3.07$ & 405 \\
6.1 & $13.5 \pm 2.51$ & $3.38 \pm 1.98$ & 41 \\
    & $13.8 \pm 2.91$ & $4.83 \pm 2.17$ & 387 \\
6.2 & $11.8 \pm 2.16$ & $1.87 \pm 1.60$ & 32 \\
    & $11.6 \pm 2.63$ & $3.23 \pm 1.76$ & 434 \\
\hline
\end{tabular}
\addtolength{\tabcolsep}{-4pt}
\label{tab:tau_stats}
\end{table}

\begin{figure}
    \centering
    \includegraphics[width=\linewidth]{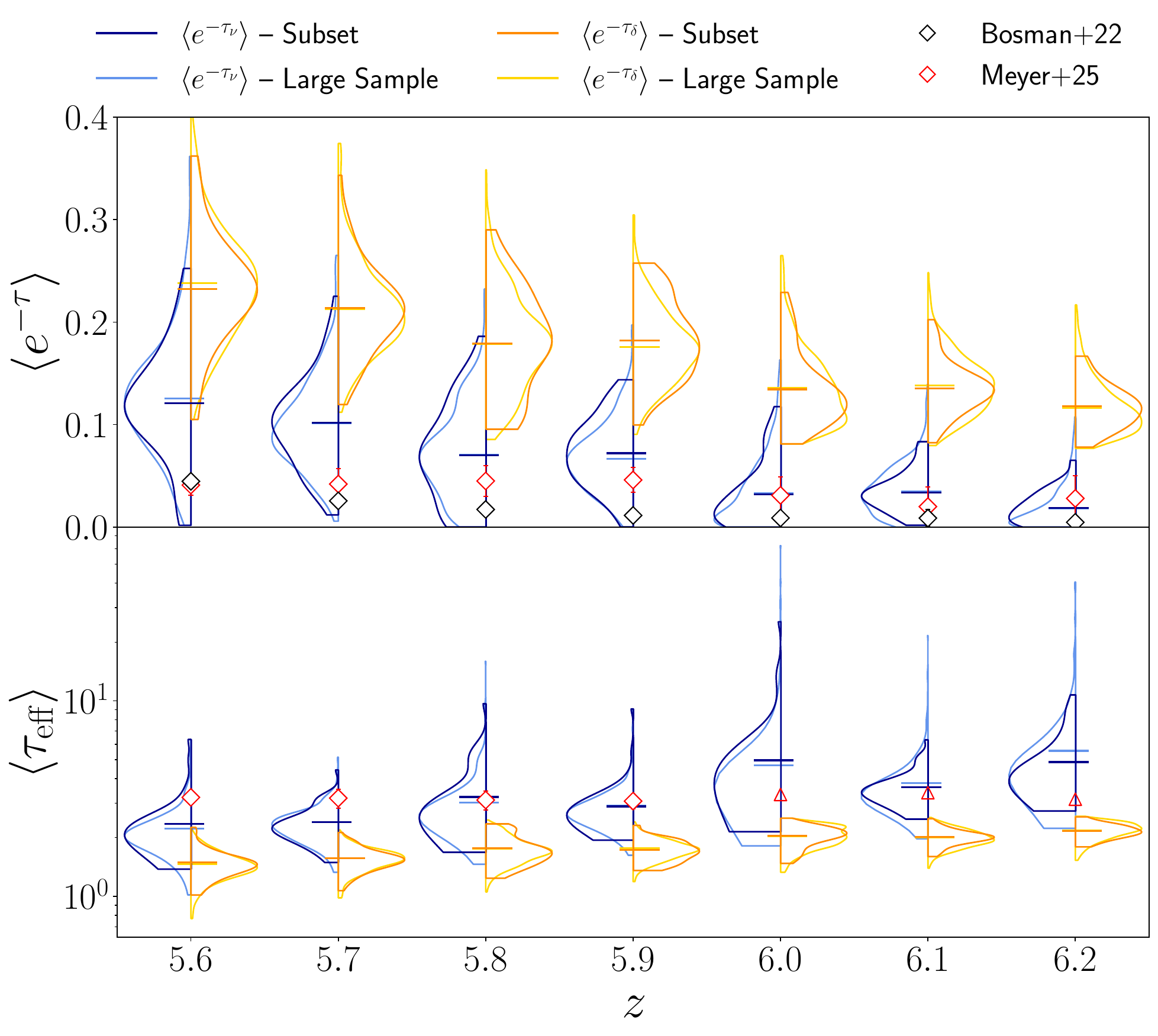}
    \caption{Violin plots of the transmission $\langle e^{-\tau} \rangle$ at different redshifts for a random sample of sightlines in the \thesan light-cone, assuming background quasar sources with redshifts varying between $5.529 < z < 6.542$. Each side shows two distributions, a smaller sample size of $300$ sightlines in dark blue/orange and a larger sample size of $3000$ sightlines in light blue/orange, representing current and future datasets. The left and right sides of each violin illustrate the systematic differences in employing Voigt ($\varphi$) and Dirac delta ($\delta$) absorption profiles, corresponding to $\langle e^{-\tau_\nu}\rangle$ and $\langle e^{-\tau_\delta} \rangle$, respectively. The $\delta$ model consistently results in much higher transmission, likely due to the damping-wing behavior being more accurately modeled by the Voigt Profile.}
    \label{fig:quasar-sightlines}
\end{figure}

\begin{figure*}
    \centering
    \includegraphics[width=0.8\textwidth]{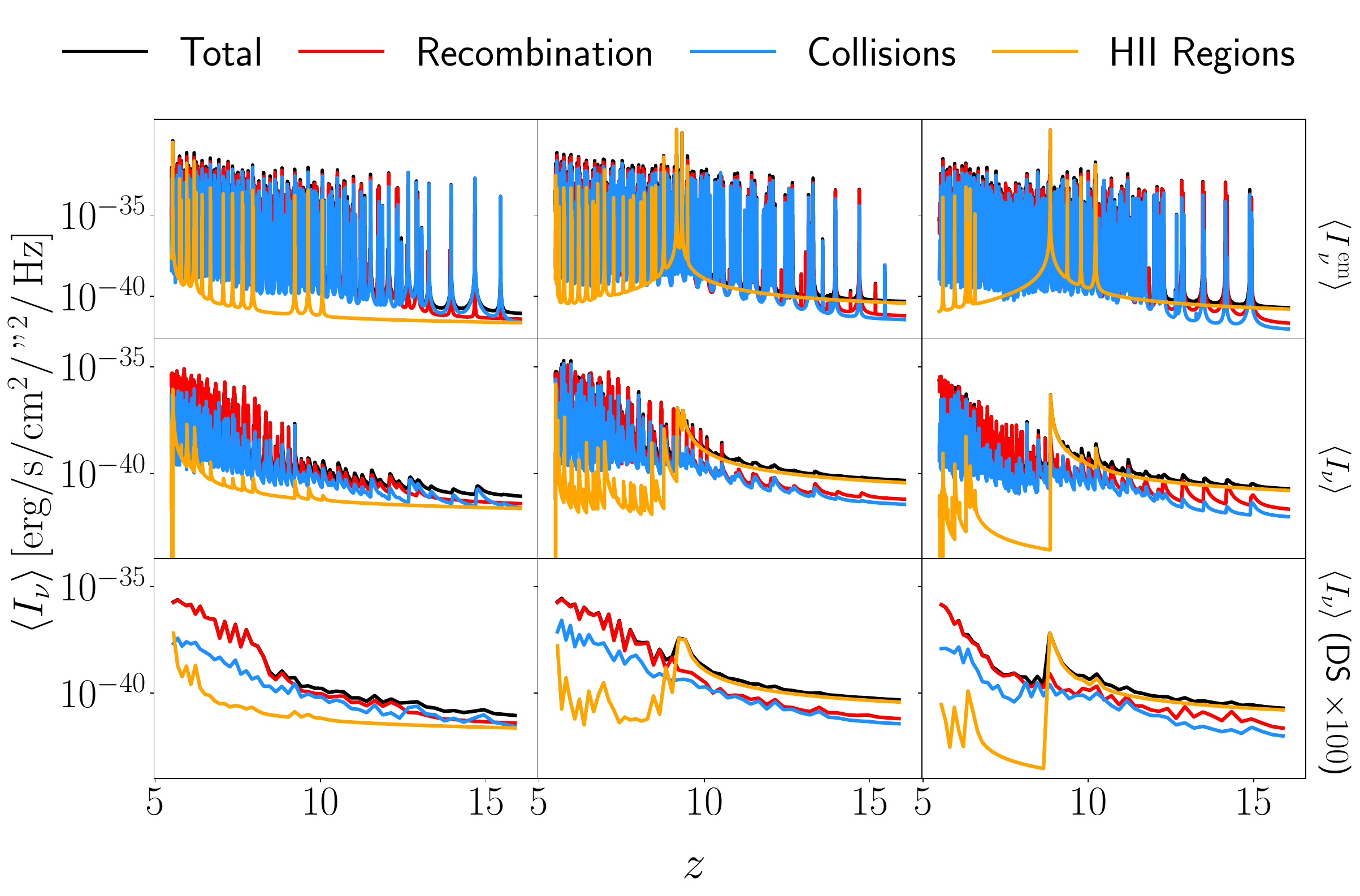}
    \caption{The top panels show the resulting Ly$\alpha$ emission-only spectral intensity for each emission mechanism for three representative pixels in our $1024\times1024$ grid. The contributions from recombination, collisional excitation, and emission from unresolved \HII regions are plotted in red, blue, and yellow, respectively, as well as the total spectral intensity in black. The middle panels show the resulting Ly$\alpha$ spectral intensity for the same three sightlines when including both emission and absorption sources. The bottom panel show the same Ly$\alpha$ spectral intensity as the middle panel, but downsampled by a factor of 100 (this sampling rate corresponds to $R\approx 52$, which is most similar to SPHEREx's $R\approx35-40$ for the relevant spectral bandpasses).}
    \label{fig:Lcum-Inu-3pix}
\end{figure*}

\section{Results}
\label{sec:results}
Using the pipeline described in Section~\ref{sec:methods}, we have generated high-resolution $1024\times1024$ grids of Ly$\alpha$ spectra, each adopting different emission/absorption profiles, for analysis. Please note that, unless otherwise stated, we use the spectra resulting from a band-integrated Voigt profile assumption. We present our results into four parts: first, we consider the absorption-only case in Section~\ref{subsec:res-absonly}; next, we include both absorption and emission in Section~\ref{subsec:res-absnem}; then, we examine a further processed data-product that mirrors LIM observations in Section~\ref{subsec:res-LIM}, and finally, we consider the impact of assuming outflows for small-scale sources in Section~\ref{subsec:outflows}.

\subsection{Absorption Only}
\label{subsec:res-absonly}
To start exploring some of the global properties of the integrated Ly$\alpha$ spectra, we follow the equations listed in the ``absorption only'' section of Table~\ref{tab:integration_schemes} to compute the Ly$\alpha$ damping wing imprint at different source redshifts. This damping wing arises due to the large cross-section of the Ly$\alpha$ line, leading to significant absorption---even in the wings of the profile---when neutral hydrogen is present. Therefore, the strength of the damping wing absorption serves as a sensitive indicator of the amount of neutral hydrogen along the LoS near the Ly$\alpha$ source. The effect has been extensively studied as a diagnostic tool for the ionization state of the IGM during the EoR \citep[e.g.][]{Miralda-Escude1998,Dijkstra2011,Mesinger2015,Mason2020,Park2021}.

To assess whether the sightline-by-sightline and statistical properties of the damping wing profiles produced by our model are reasonable, we can compare the damping wing profiles for different source redshifts.  In the top panel of Figure~\ref{fig:dampingwing_combined}, we plot the transmission of a random sightline using a Voigt profile for absorption (i.e., an individual pixel in our $1024\times1024$ grid). Note that, due to the nature of the produced light slabs, all sightlines run parallel to the line of sight and do not originate from the halo center, hence explicitly probing the IGM effects. As expected, the strength of the damping wing absorption diminishes significantly as we go to lower source redshifts, and correspondingly, lower global neutral hydrogen fractions as reionization progresses. At $z_{\rm{source}} \gtrsim 8$ the IGM is predominantly neutral, yielding saturated absorption blue-ward of the Ly$\alpha$ line center and broad red-wing absorption. Conversely, at $z_{\rm{source}}=6$, the Universe is largely ionized, and there are significant transmission peaks blueward of line-center due to ionized regions along the line of sight.

\begin{figure}
    \centering
    \includegraphics[width=1\columnwidth]{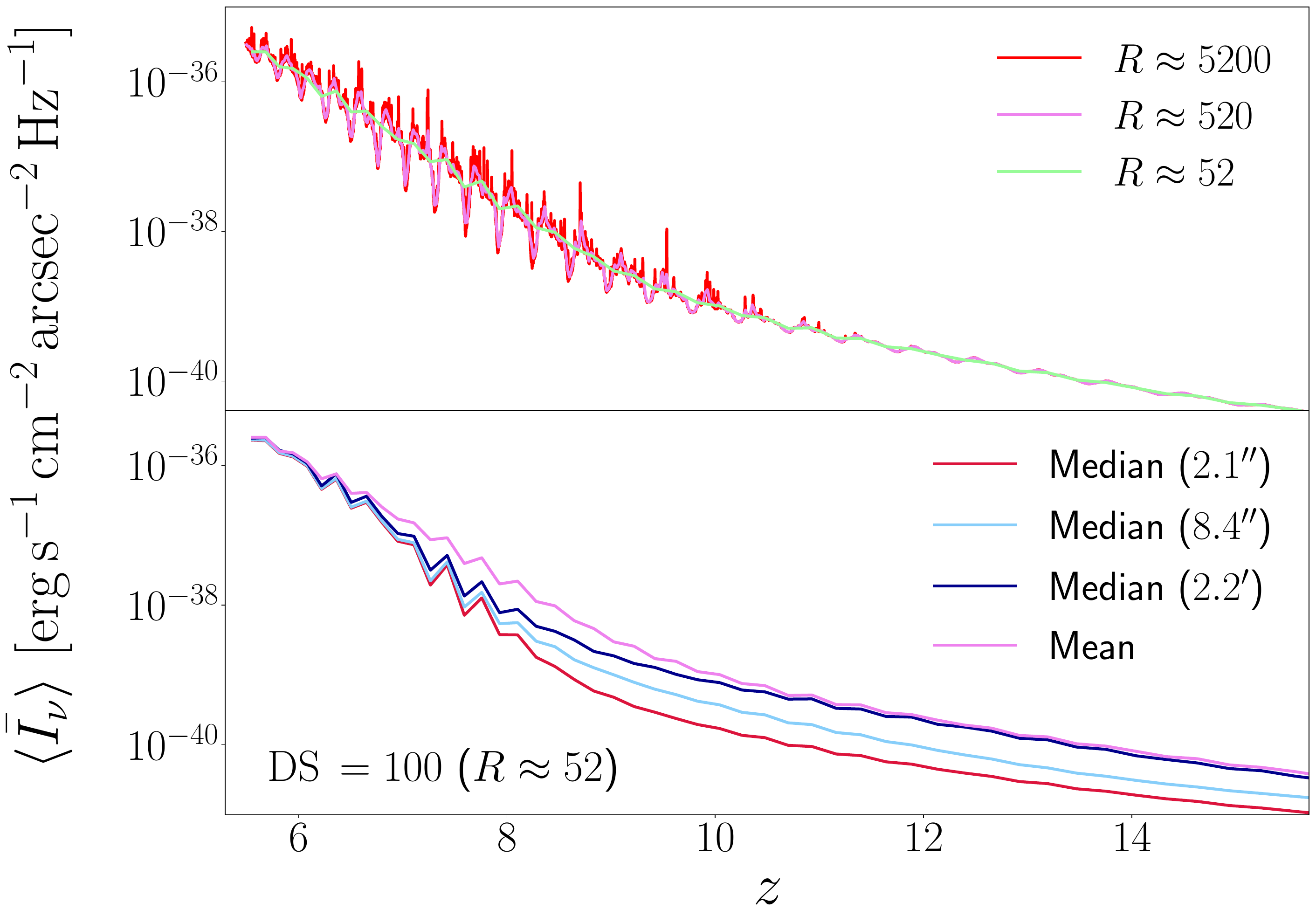}
    \caption{The top panel shows the resulting mean of our total spectral intensity $\langle I_\nu \rangle$ for different spectral resolutions of our grid. Our nominal sampling rate $n=5000$ results in a spectral resolution $R\approx 5200$, plotted in red. Downsamplings of $10\times$ and $100\times$ are shown in violet and pale green respectively. In the bottom panel, we plot the resulting statistics of different spatial resolutions of our grid at a constant spectral resolution $R\approx 52$; the mean is plotted in violet, along with the median for different spatial downsamplings of our $1024\times1024$ grid for comparison. These downsamplings correspond to pixel scales at $z=5.5$ of $2.1''$ (the full-resolution $1024\times1024$ grid), $8.4''$, and $2.2'$, shown as red, light blue, and dark blue lines respectively. We can see that the mean is biased high but is most representative of signals extracted from Ly$\alpha$ LIM methods.}
    \label{fig:total_intensity}
\end{figure}

We validate that the statistical properties of the damping wing profiles produced by our model are reasonable in the bottom panel of Figure~\ref{fig:dampingwing_combined}, where we calculate the median and $1\sigma$ deviation in transmission curves over all pixels in our $1024\times1024$ grid. For pre-reionization redshifts, there is very little scatter red-ward of line center, reflecting the relatively more uniform IGM conditions. In contrast, by $z_{\rm{source}}=6$, there is significantly more variation across sightlines in both the red-ward absorption and blue transmission spikes, which is expected during the later stages of patchy reionization. This is particularly the case as ionized bubbles begin to overlap, and the distribution of neutral hydrogen varies across different regions \citep{McQuinn2007, Dijkstra2011}. These results are qualitatively consistent with observational studies of high-redshift quasars and galaxies, with significant variation across pixel sightlines \citep{Fan_2006, Becker2015}.

To directly compare our results to observational measurements of the mean Ly$\alpha$ transmission for $4.75 < z < 6.25$, we average over intervals of $\Delta z=0.1$ as done in Table~4 of \citet{BoDa22} and assume that the source redshifts for the background quasars are randomly sampled across a uniform range $5.529 < z < 6.542$, adopting two sample sizes: one similar to the size of the observational sample in \citet{BoDa22}, and another with 3000 sightlines total to ensure convergence. The resulting average transmissions are listed in Table~\ref{tab:tau_stats} along with the $1\sigma$ standard deviation and the number of sightlines with data covering that redshift range for each sample distribution. The distribution of transmission at various redshifts is shown in Figure~\ref{fig:quasar-sightlines}. Overall, \thesan agrees well with the observational data of \citet{Meyer_2025}, shown as red diamonds, but produces too much transmission at lower redshifts, which we attribute to a slightly too early timing of reionization \citep[see][]{Garaldi2022, Garaldi2024a, Garaldi2024b}. We nevertheless emphasize the importance of modeling absorption accurately with a Voigt rather than a Dirac delta profile, and we consider our band-integrated continuous Doppler-shifting formalism an improvement for such applications, having clearly produced more realistic transmission distributions as a function of redshift in comparison to the Dirac delta profile when compared to observations.

\begin{figure}
    \centering
    \includegraphics[width=\columnwidth]{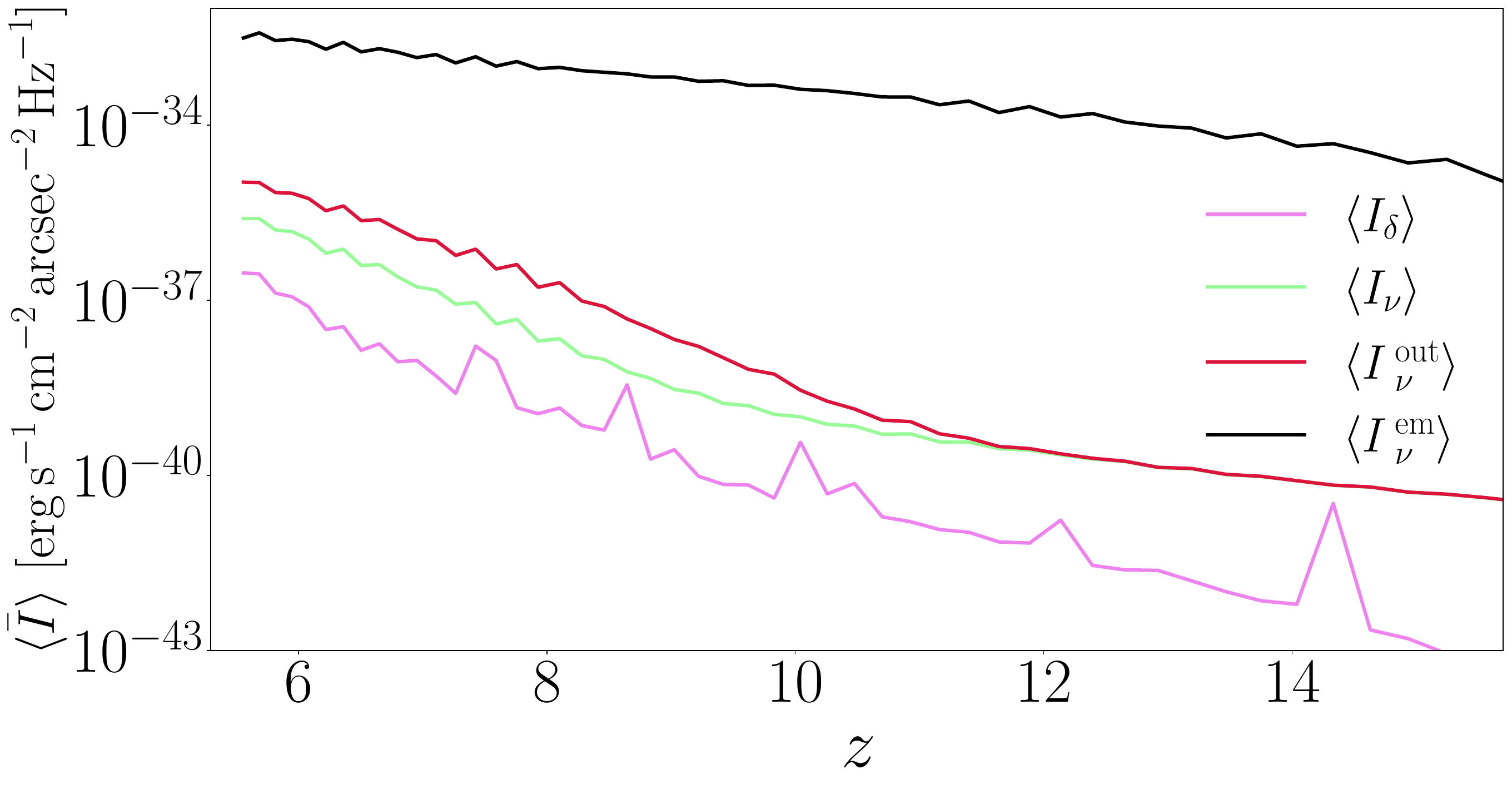}
    \caption{Mean spectral intensities for different analytic forms of our emission/absorption profiles at a fiducial spectral resolution $R\approx52$. Here we show statistics for the band-integrated Dirac delta $\langle I_{\delta}\rangle$ in violet, the band-integrated Voigt $\langle I_{\nu}\rangle$ in light green, the band-integrated Voigt with no self-absorption $\langle I\,_{\nu}^{\rm{out}}\rangle$ in crimson, and the band-integrated emission-only Voigt $\langle I\,_{\nu}^{\rm{em}}\rangle$ in black. The emission-only signal (black) is clearly significantly suppressed with the inclusion of absorption (light green), with our outflow model allowing for some Ly$\alpha$ escape at low redshift due to there being no self-absorption at the cell level. The absorption+emission signal $\langle I_{\nu} \rangle$ is clearly a pessimistic lower limit, emphasizing the importance of scattering.}
    \label{fig:intensity_intmethods}
\end{figure}

\begin{figure}
    \centering
    \includegraphics[width=\columnwidth]{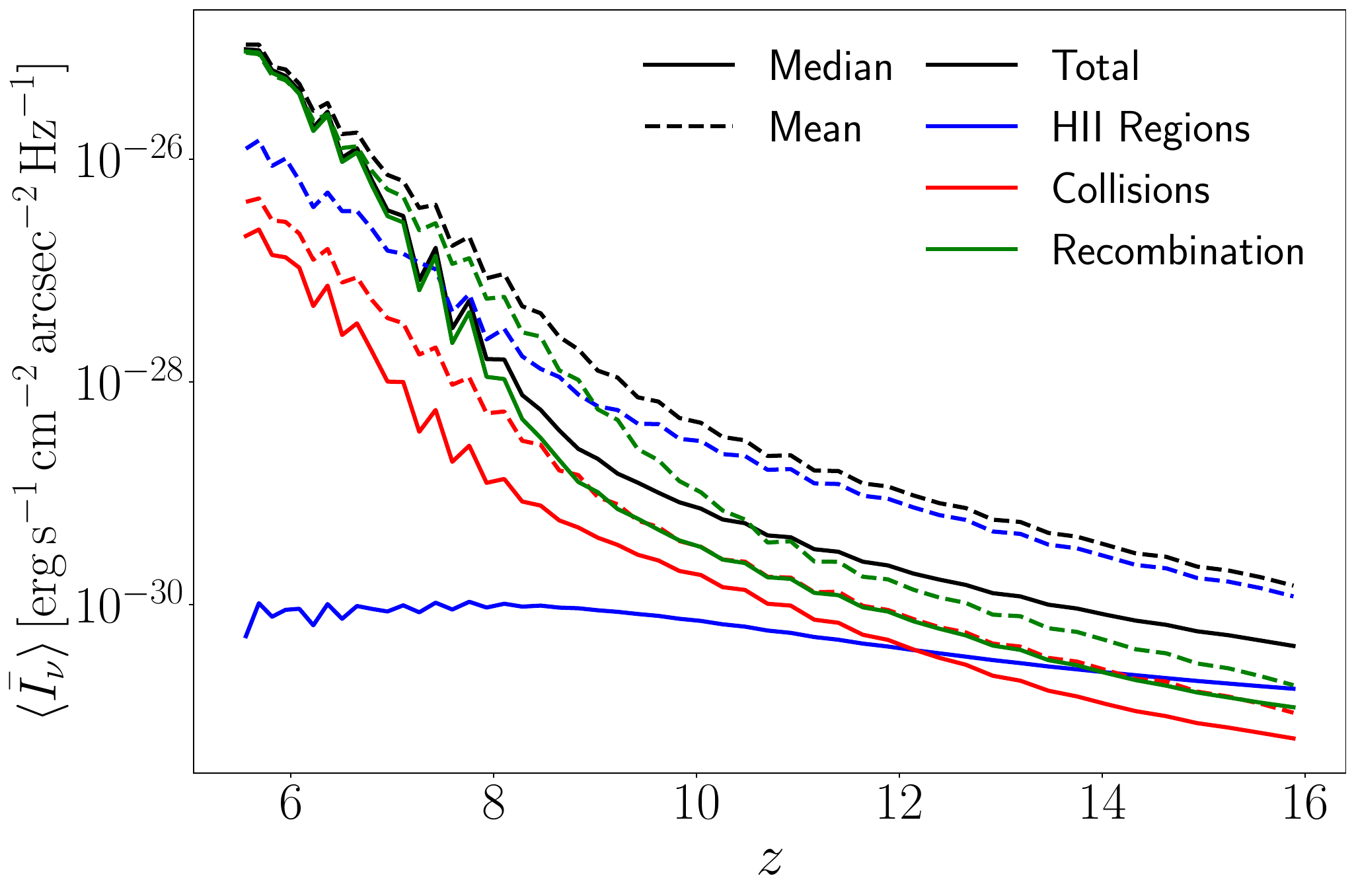}
    \caption{The mean and median results for the specific intensity as a function of redshift are plotted for each emission mechanism at a fiducial spectral resolution $R\approx52$. Median and mean contributions from each emission mechanism are plotted in solid and dashed lines respectively, with the unresolved \HII regions, collisional excitation, and recombination shown as blue, red, and green lines respectively, with the total contribution plotted as a solid black line. The mean contribution from \HII regions starts to be dominated by recombination for $z\lesssim 9$ due to significant self-absorption (See Figure~\ref{fig:compare-outflow-pixels}).}
    \label{fig:med_intensity_1024}
\end{figure}
\begin{figure*}
    \centering
    \includegraphics[width=0.9\textwidth]{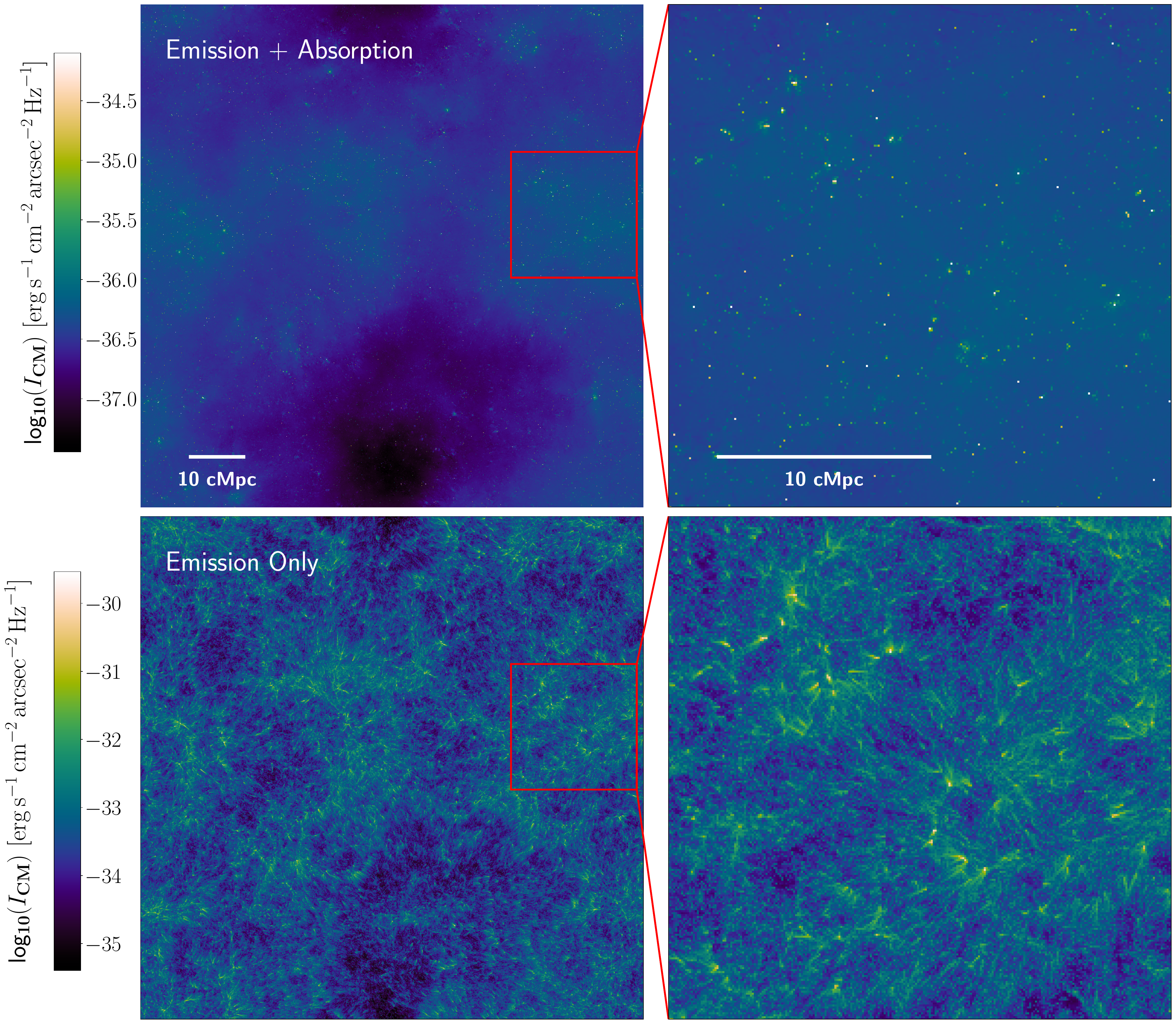}
    \caption{The resulting spectral intensity map when integrating over the entire redshift range of our simulation, $z
    \in [5.49,16.05]$, with $\langle I_{\nu} \rangle$ shown on the top panel and $\langle I\,_{\nu}^{\rm em} \rangle$ shown on the bottom panel. We show the full 95.5 cMpc box on the left, and a zoom-in of a 24$\times24$ cMpc$^{2}$ square patch of the image--indicated by the red box--on the right.}
    \label{fig:channel_maps_inset_1024}
\end{figure*}
\label{subsec:res-LIM}
\begin{figure*}
    \centering
    \includegraphics[width=0.9\textwidth]{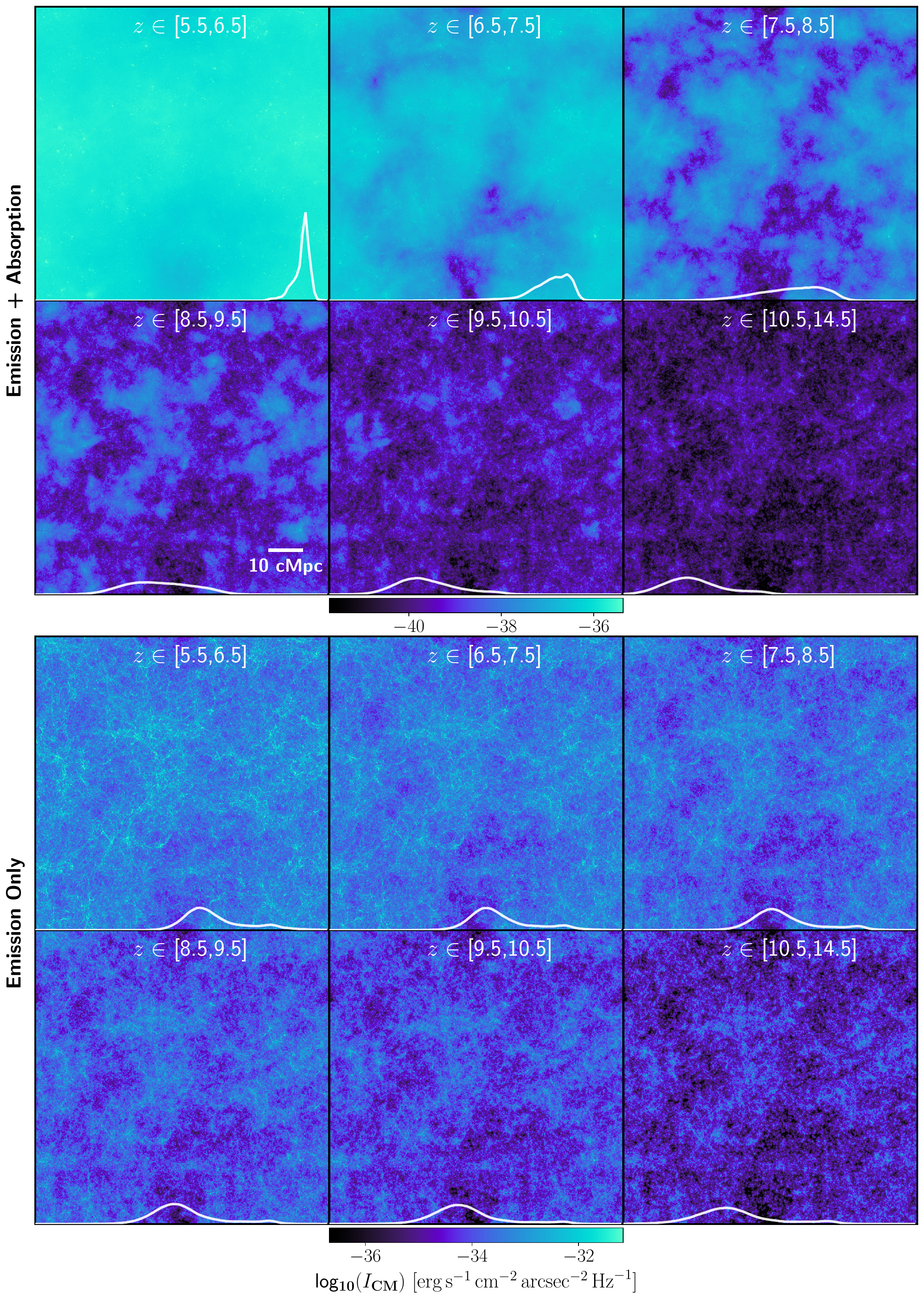}
    \caption{
    Channel maps when integrating over respective redshift ranges [$z_{i}$,$z_{f}$] indicated in the title of each subplot, both when employing absorption+emission (top six panels) and emission-only (bottom six panels) formalisms. The histograms of the $I_{\rm{CM}}$ distributions also plotted in white at the bottom of each subplot, with the axis indicated by the labels on the colorbar at the bottom center of the figure.}
    \label{fig:6-channel-maps}
\end{figure*}
\subsection{Absorption and Emission}
\label{subsec:res-absnem}
We now employ the full RT equation to explore the combined effects of absorption and emission on the global properties of Ly$\alpha$ transmission. To illustrate the diverse behavior along different sightlines, we select three representative sightlines from our resulting $1024\times1024$ spectral grid. For the example sightlines in Figure~\ref{fig:Lcum-Inu-3pix}, each occupying one column, we plot the Ly$\alpha$ spectral intensity as a function of redshift when including only emission sources (top row) vs when including both emission and absorption sources (middle row); we also plot the spectrally downsampled ($\times 100$) version of the second row in the last row, which corresponds to $R\approx52$ (this sampling rate is most similar to SPHEREx, which has a spectral resolution $R\approx35-40$ for the relevant spectral bandpasses).

We can quickly notice a few differences between the sightlines we have plotted; for example, it is clear that we cross a bright star-forming region at $z\sim9$ in the unresolved \HII region emission for the second and third columns (plotted in yellow), hence significantly contributing to the total emission. Therefore, emission is sensitive to the stochastic proximity to star-forming galaxies, and the reprocessing of the signal as a result of local scattering and effective absorption at larger distances is sensitive to both self-shielding and reionization.

We plot the mean total spectral intensity produced by our fiducal $\langle I_{\nu}\rangle$ model for different spectral resolutions in the top panel Figure~\ref{fig:total_intensity}, with $R\in[5200,520,52]$ represented as red, violet and light green lines respectively. We can see a steady logarithmic increase in the mean intensity as a function of redshift for $z\lesssim 10$. In the bottom panel, we plot the same mean intensity in violet, along with the median intensity for different spatial resolutions of our grid. The highest resolution corresponds to a pixel scale of $2.1''$ and is shown in red, with downsamplings $\times 4$ and $\times 64$ shown in light blue and dark purple respectively. All of the lines in the bottom panel are shown for our fiducial spectral resolution of $R\approx 52$. he LIM method aims to observe the EoR at low spatial resolution of $\sim 6''$, which is most closely represented by our grid downsampled to  $8.4''$. Note that the trend in the median spectral intensity clearly approaches the mean as we increase the coarse-graining factor; however, the mean is the most representative metric for LIM methods (even more representative are the power spectra of the spatial fluctuations, discussed in Section~\ref{subsec:res-LIM}). We can see that at intermediate redshifts, the two metrics noticeably diverge, with the mean tracking the more overdense regions of the Universe, and the highest-resolution median tracking average contributions from both the fainter and the more concentrated emission. As we approach the lowest redshift in the simulation, the plateau in intensity across all metrics likely signals the end of the EoR. 

In Figure~\ref{fig:intensity_intmethods}, we plot the mean spectral intensities for various band-integrated functional forms of our emission/absorption; we show results when including both emission and absorption in the radiative transfer, with $\langle I_{\nu} \rangle$, $\langle I_{\delta} \rangle$, and $\langle I\,_{\nu}^{\rm out} \rangle$ shown in light green, red, and violet respectively, as well as when only including emission and using our fiducial band-integrated Voigt profile $\langle I\,_\nu^{\rm em} \rangle$ in black. We can see that our emission-only signal is much brighter at higher redshifts, with the strong neutral hydrogen absorption overwhelming any possible Ly$\alpha$ emission. At lower redshift, we can see deviation between our $\langle I_{\nu} \rangle$ and $\langle I\,_{\nu}^{\rm out} \rangle$; this result is expected, with the latter model effectively not including same-cell (``local'') self-absorption when closely inspecting Eq.~(\ref{eq:voigt-band-noSA}), hence allowing for emission that is redward enough of line-center to slip through as the Universe becomes more ionized. This line does not match the emission-only signal likely due to  strong absorption even in the wings of the profile whenever caught in a neutral region (i.e., it is caught in the succeeding cells), which is more probable in the overdense regions where we have stronger emission. The strong suppression of the signal due to absorption is worth exploring further and is discussed in more detail in Sections~\ref{subsec:res-LIM} and especially~\ref{subsec:outflows}.

Finally, we plot the median and mean across all sightlines of the different contributions to the total spectral intensity from each emission mechanism in Figure~\ref{fig:med_intensity_1024}. It is again interesting to note that the contribution from stars plateaus and decreases after $z\sim9$, while the contribution from recombination is most dominant. The decrease in contribution from stars, as was also seen in the second and third columns of Figure~\ref{fig:Lcum-Inu-3pix}, is likely due to the presence of self-shielding in more massive galaxies that prevents the Ly$\alpha$ photons from escaping without help from resonant scattering and, perhaps secondly, decreased star formation in dwarf galaxies that are quenched during reionization. The mean contribution from stars also comes second to that from recombinations for $z<9$. This trend points towards the importance of including the effects of resonant Ly$\alpha$ scattering back into the LoS, and the presence of significant self-absorption is confirmed in Section~\ref{subsec:outflows}; the inclusion of scattering effects in the radiative transfer will be addressed in future work.

\subsection{Predictions for LIM}

\begin{figure}
    \centering
    \includegraphics[width=\columnwidth]{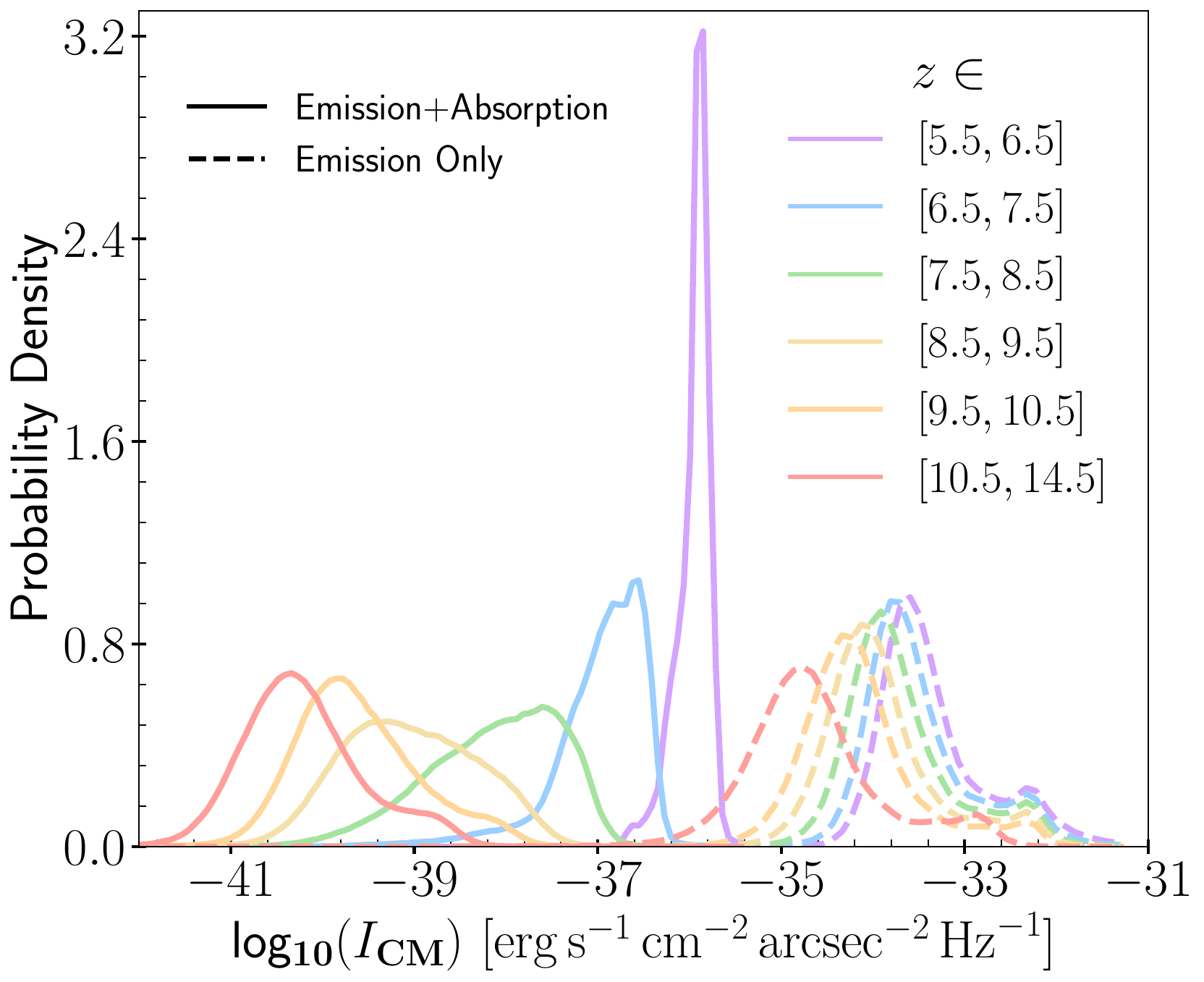}
    \caption{Channel map intensity distribution across all pixels in the $1024\times1024$ image, for a variety of different redshift bands. These distributions are plotted both for the emission+absorption formalism (solid lines) and for emission only (dashed lines). Note that the $y$-axis is showing the probability density (i.e., it is normalized for each distribution).}
    \label{fig:histograms}
\end{figure}
\begin{figure*}
    \centering
    \includegraphics[width=\textwidth]{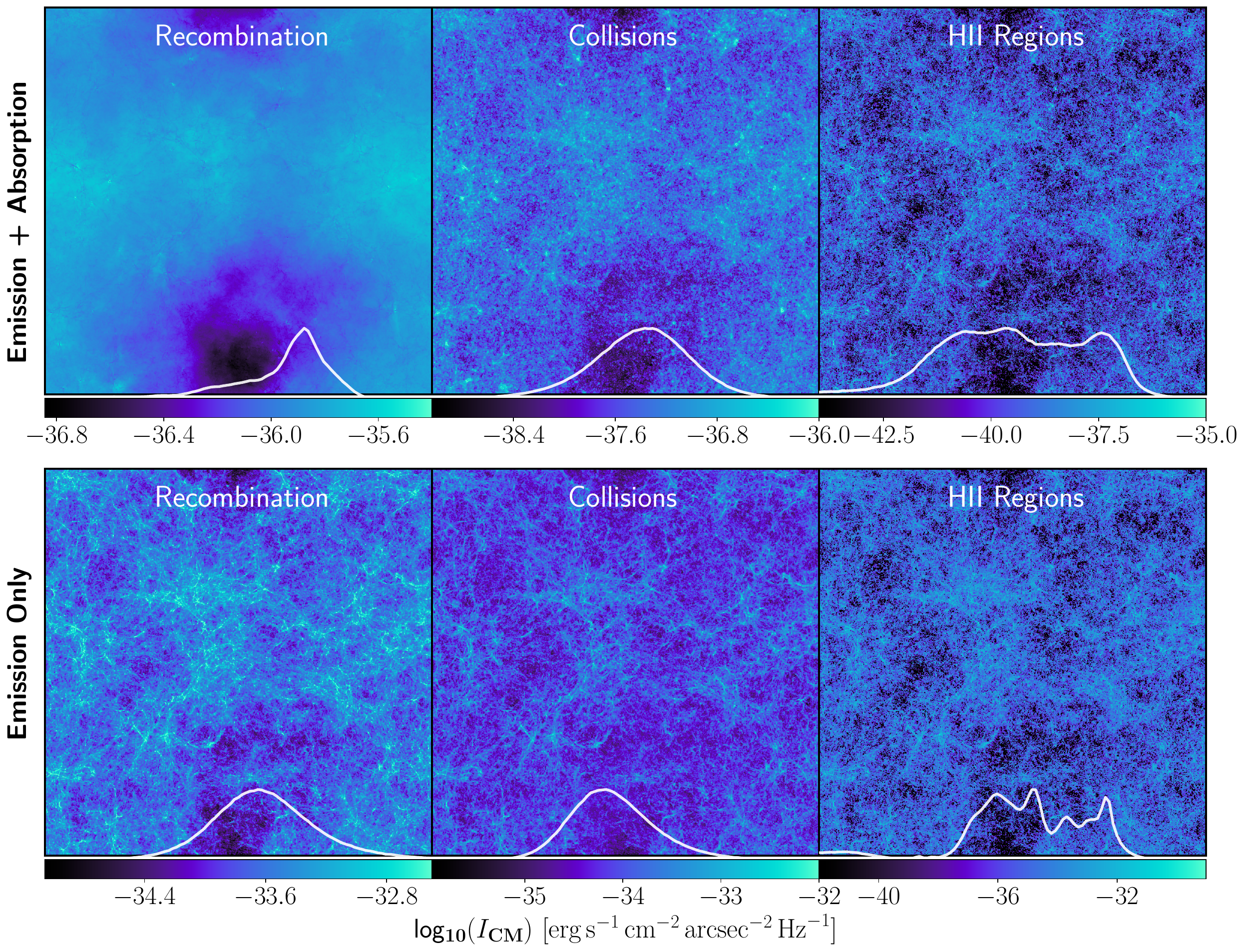}
    \caption{Ly$\alpha$ channel maps for the redshift band $z\in[5.5,6.5]$ broken down into the different emission mechanisms, with the radiative transfer including emission and absorption (using $\langle I_{\nu}\rangle$) on the top panel, and emission only in the bottom panel (using $\langle I_{\nu}^{\rm{em}}\rangle$). The left panel shows recombination emission, the center panel shows emission from collisional excitation, and the right panel shows unresolved emission from \HII regions.}
    \label{fig:channel_maps_emissionmech}
\end{figure*}

To start probing the evolution in spatial fluctuations in our Ly$\alpha$ intensity in a manner similar to what we will see in LIM experiments, we compute Ly$\alpha$ ``channel maps'', henceforth denoted as $I_{\rm{CM}}$, integrating over some range of frequencies $[\nu_1,\nu_2]$:
\begin{equation}
    I_{\rm{CM}} = \frac{\int_{\nu_1}^{\nu_2} I_\nu\,\text{d}\nu}{\int_{\nu_1}^{\nu_2} \text{d}\nu} \, ,
\end{equation}
where the frequency range is determined by the redshift range of interest. To start, we obtain a spatial map integrating from the start to end redshifts of our simulation $z\in[5.49,16.05]$, shown in Figure~\ref{fig:channel_maps_inset_1024} for both emission and absorption ($\langle I_{\nu}\rangle$, top panel), and emission only ($\langle I\,_{\nu}^{\rm em}\rangle$, top panel) implementations. The left panel shows the zoomed out view for our 95.5 cMpc box, and the right panel shows a zoom-in of a $\sim 24\times24$ cMpc$^{2}$ square patch of the image. The inset highlights the inclusion of more concentrated emission from overdense regions as well as the more diffuse emission that contributes to reionization across time and space--a key contribution that is, by design, expected from LIM observations. 

To gain more physical intuition for how the Ly$\alpha$ line intensity is spatially evolving, we can start to choose the spectral/redshift bands for our channel maps. In Figure~\ref{fig:6-channel-maps}, we show channel maps for six different redshift bands across our spectra for the full $1024\times1024$ grid, both in the form of emission+absorption (top 6 panels) and emission only (bottom 6 panels) using our fiducial Voigt models. In the bottom section of each subplot, we also plot the distribution in spectral intensity for the corresponding band, with the axis indicated by the colour bar at the bottom center of the plot. We notably have much harder and more strongly-peaked spectral intensity distributions as we go to lower redshift bands in our emission+absorption model; the onset and progression of patchy reionization is apparent, with the more distinct ionized blobs in the $z\in[7.5,8.5]$ panel starting to merge in the lead-up to rapid reionization at $z\sim5.5-7.5$. This trend is more distinguishable in Figure~\ref{fig:histograms}, which explicitly shows the $I_{\rm{CM}}$ distribution in six different redshift bands. There is an especially prominent transition in the strength of the peak as the Universe becomes significantly reionized from $z\in[6.5,7.5]$ to $z\in[5.5,6.5]$, with very little overlap except in the wings of the strongly peaked $z\in[5.5,6.5]$ distribution. We can see in our emission-only panels and histograms that this strong peak is absent, indicating that it arises from absorption at the locations where Ly$\alpha$ emission is produced. Because most of the highly emitting regions sit in overdense environments that retain comparatively high neutral hydrogen columns, they experience substantial self-absorption, naturally producing the strong, narrow peak in the absorption-included case. Including scattering would likely lessen these effects—consistent with the significant suppression of our signal (see Section~\ref{subsec:outflows} for validation in the \HII-region case). Nonetheless, some of the absorption is likely still physical, reflecting uncorrelated scattering at large distances for frequencies blueward of line center being converted to diffuse background light.

As expected, these differences also appear when we examine the emission mechanisms separately in Figure~\ref{fig:channel_maps_emissionmech}, where we plot channel maps for the range $z\in[5.5,6.5]$ only, both for emission and absorption (top panel) and emission only (bottom panel). Recombination emission appears to be more diffuse in its spatial distribution, with its spectral intensity distribution developing a strong peak once absorption is included, consistent with its association with the overdense regions that dominate the total signal. The collisional excitation and \HII-region components show a similarly strong overall suppression in spectral intensity when including absorption.

\begin{figure*}
    \centering
    \includegraphics[width=0.8\textwidth]{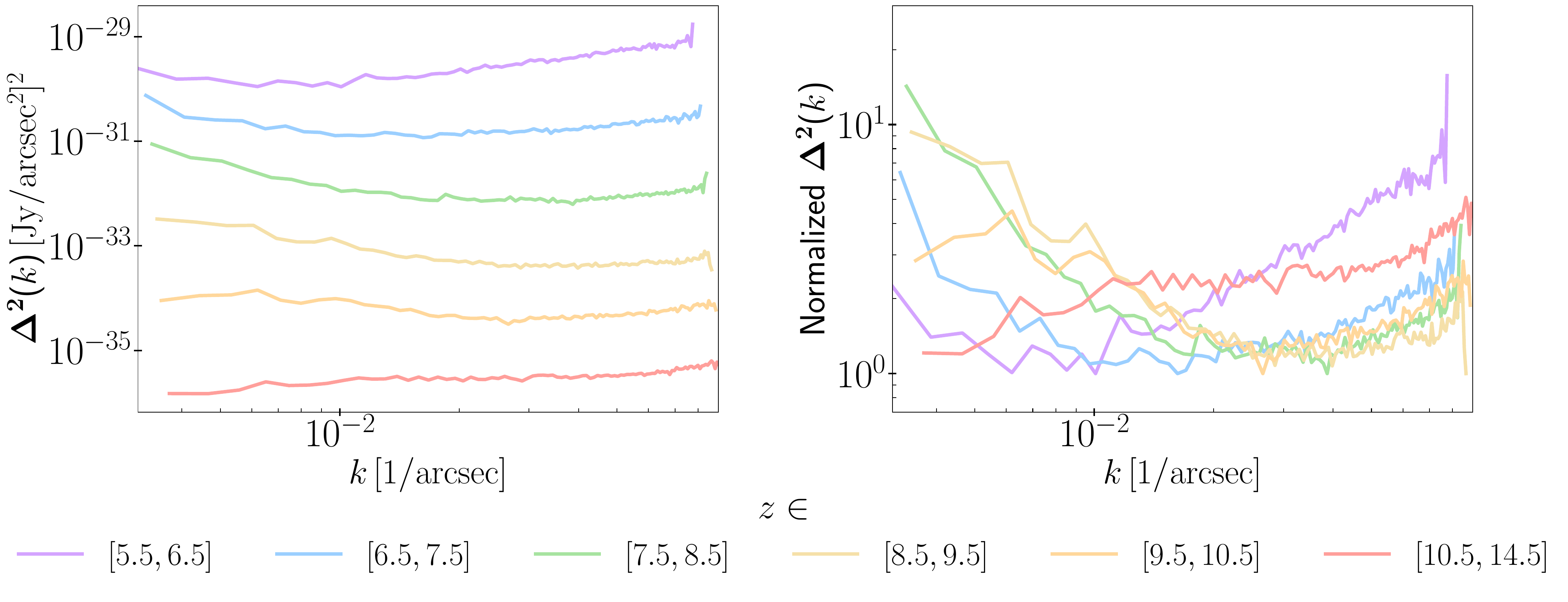}
    \caption{2D power spectra are plotted as a function of wave number $k$ for channel map integrations performed over a variety of different redshift bands. On the left panel we plot the absolute power spectra, and on the right panel we plot the power spectra normalized such that there is a common minimum between all of the curves. The SPHEREx Deep Field's limits (based on projected surface brightness noise levels) are also plotted as black triangles for the band corresponding to Ly$\alpha$ at $z\in[5.5,6.5]$. Spectra are shown for a downsampling of the original grid to an angular resolution of 8.4''.}
    \label{fig:power-spectra-normalized}
\end{figure*}

In Figure~\ref{fig:power-spectra-normalized}, we present the power spectra of our channel maps across different redshift ranges, which serve as a key metric for the evolution of reionization at all angular scales. To mitigate the appearance of Poisson noise at higher $k$ values--manifesting as a $k^2$ slope---we downsampled our $1024 \times 1024$ grid to a coarser $256 \times 256$ resolution, corresponding to a pixel scale of $8.4''$ at $z=5.5$. The power spectrum for the highest redshift band, $z \in [10.5, 14.5]$, is predominantly concentrated at the smallest angular scales. As we move to lower redshifts, the power gradually shifts towards larger angular scales, marking the onset of reionization as the size of the ionized bubbles grow and merge; this progression culminates in a pronounced peak at the largest scales for $z \in [6.5, 7.5]$. We also notice that the slope of the power spectrum at $k\gtrsim 10^{-2}\,\mathrm{arcsec}^{-1}$ becomes steeper with the progress of reionization---a similar result to that of \citet{AmVi2025}, where the power spectrum steepens as the neutral fraction decreases. The spectrum then starts to flatten for $z \in [5.5, 6.5]$, indicating a proximity to a fully reionized universe across scales. These changes in the power spectrum hence provide valuable insights into the timing and progression of reionization. 

\begin{figure}
    \centering
    \includegraphics[width=\columnwidth]{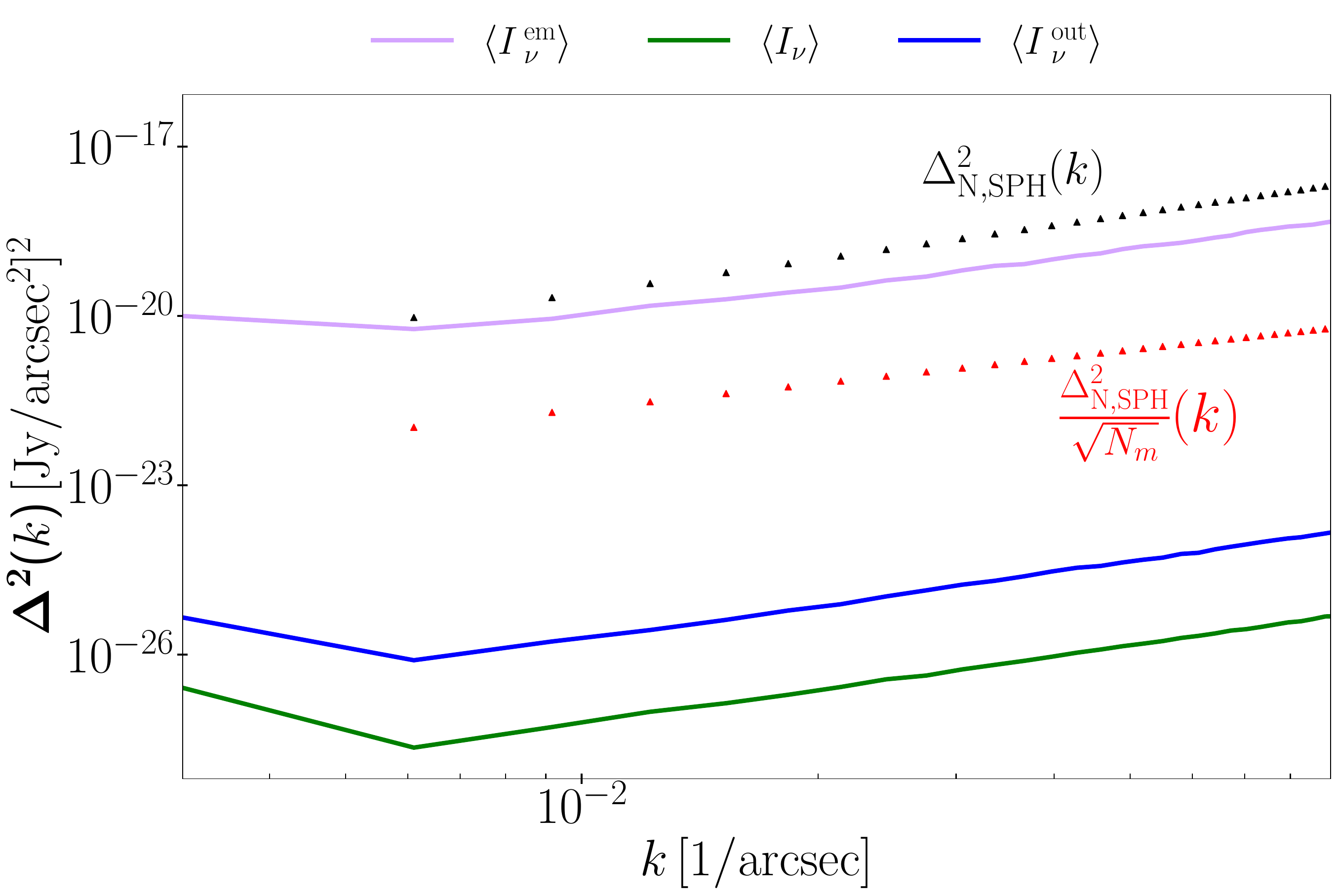}
    \caption{The 2D power spectrum of our highest resolution grid is plotted as a function of wave number $k$ for a channel map redshift range $[5.632,5.798]$ corresponding to one of SPHEREx's spectral bandpasses; these results are shown using a band-integrated Voigt, with emission-only ($\langle I\,_{\nu}^{\rm em} \rangle$) in lavender, Emission+Absorption ($\langle I_{\nu} \rangle$) in green, and our outflow model (no self-absorption at the cell level -- $\langle I\,_{\nu}^{\rm out} \rangle$) in blue. The SPHEREx Deep Field limits (based on projected surface brightness noise levels) are also plotted as black triangles, along with the same limits scaled by the number of modes per $k$-bin $N_{m}(k)$ as $1/\sqrt{N_{m}(k)}$. Our emission-only signal is detectable by SPHEREx when accounting for the number of modes in the signal-to-noise estimate, but our pessimistic emission+absorption signal is not.}
    \label{fig:SPHEREx-compare}
\end{figure}

In Figure~\ref{fig:SPHEREx-compare}, we plot the power spectra for our $\langle I_{\nu} \rangle$, $\langle I_{\nu}^{\rm em} \rangle$, and $\langle I\,_{\nu}^{\rm out} \rangle$ models in green, lavender, and blue, respectively, along with the SPHEREx Deep Field surface-brightness noise level\footnote{https://github.com/SPHEREx/Public-products} $\Delta_{\rm N,SPH}^{2}(k)$ shown as black scatter points. We also include the corresponding noise after accounting for the number of modes per $k$-bin in the $200$ deg$^{2}$ deep fields, $\Delta_{\rm N,SPH}^{2}(k)/\sqrt{N_m}$, shown in red. For this comparison we use our highest-resolution grid, which is why the power-spectrum amplitudes differ from those in Figure~\ref{fig:power-spectra-normalized}. The spectra shown correspond to channel maps covering one of SPHEREx’s bandpasses, spanning $z\in[5.632,5.798]$. We find that the emission-only signal is detectable once the number of modes per $k$-bin is included in the noise estimate, whereas the full emission+absorption signal lies below the detection threshold. Removing self-absorption at the cell level (blue curve) increases the signal by roughly an order of magnitude, again indicating the dominance of self-absorption and supporting the interpretation that $\langle I_{\nu} \rangle$ represents a lower limit on the true signal—both because scattering would alleviate the artificially strong self-absorption produced by our scattering-free treatment, and because of the Jensen’s inequality limit applied in Eq.~(\ref{eq:absem_voigt_band}).

\begin{figure*}
    \centering
    \includegraphics[width=0.9\textwidth]{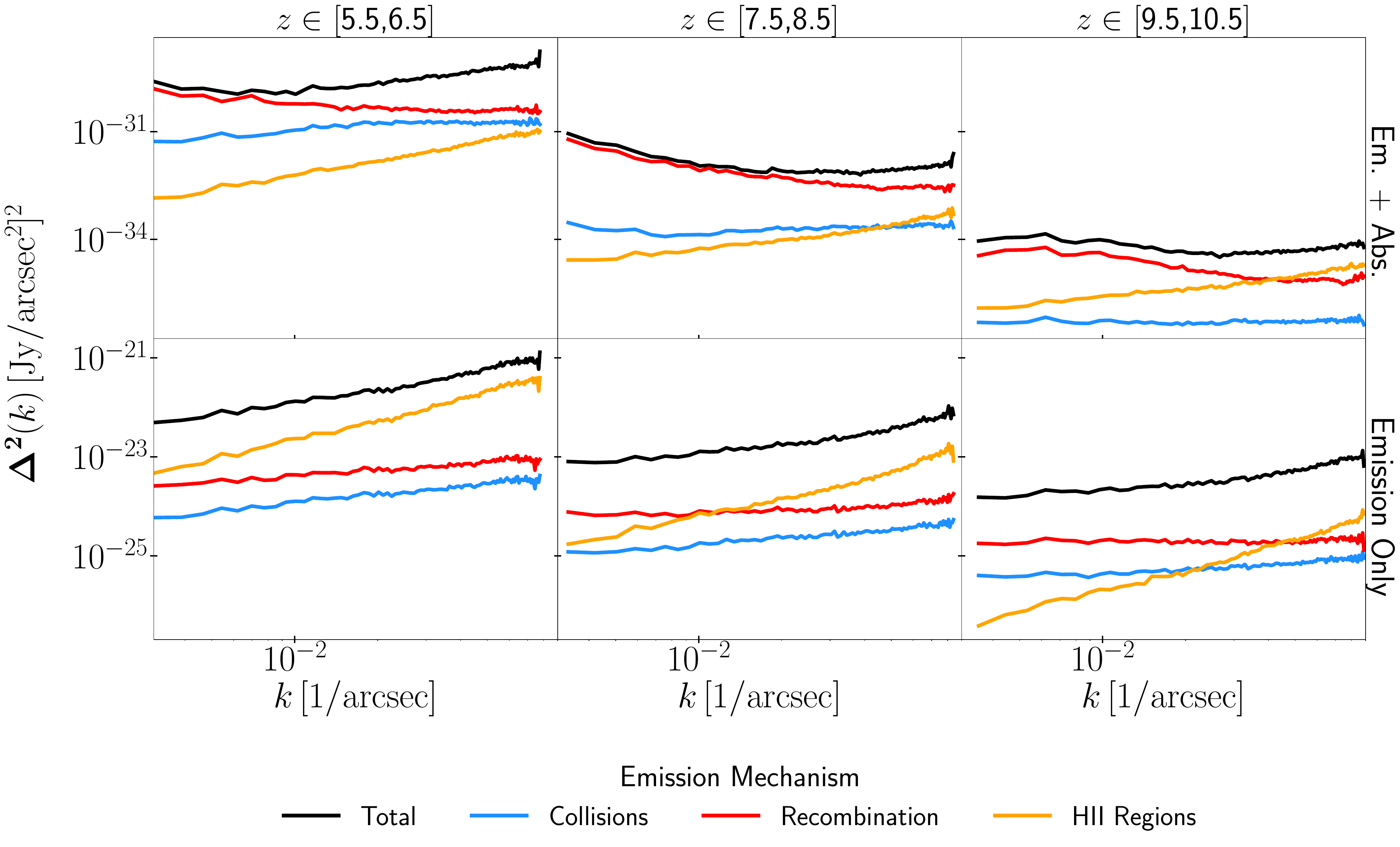}
    \caption{2D power spectra of the channel maps plotted as a function of wave number $k$ for different emission mechanisms, with each redshift/spectral band's map occupying one subplot. The resulting spectra are shown when including both emission and absorption on the top panel, and when only including emission on the bottom panel. The emission contributions from collisional excitation, recombination and \HII regions are plotted in blue, red, and yellow respectively, and the total emission is plotted in black. Spectra are shown for a downsampling of the original grid to an angular resolution of 8.4''. There is clear evolution of the spectrum shape for different redshift bands, hence indicating its sensitivity to the timing and progress of reionization.}
    \label{fig:power-spectra-emissionmech}
\end{figure*}
\begin{figure*}
    \centering
    \includegraphics[width=0.85\textwidth]{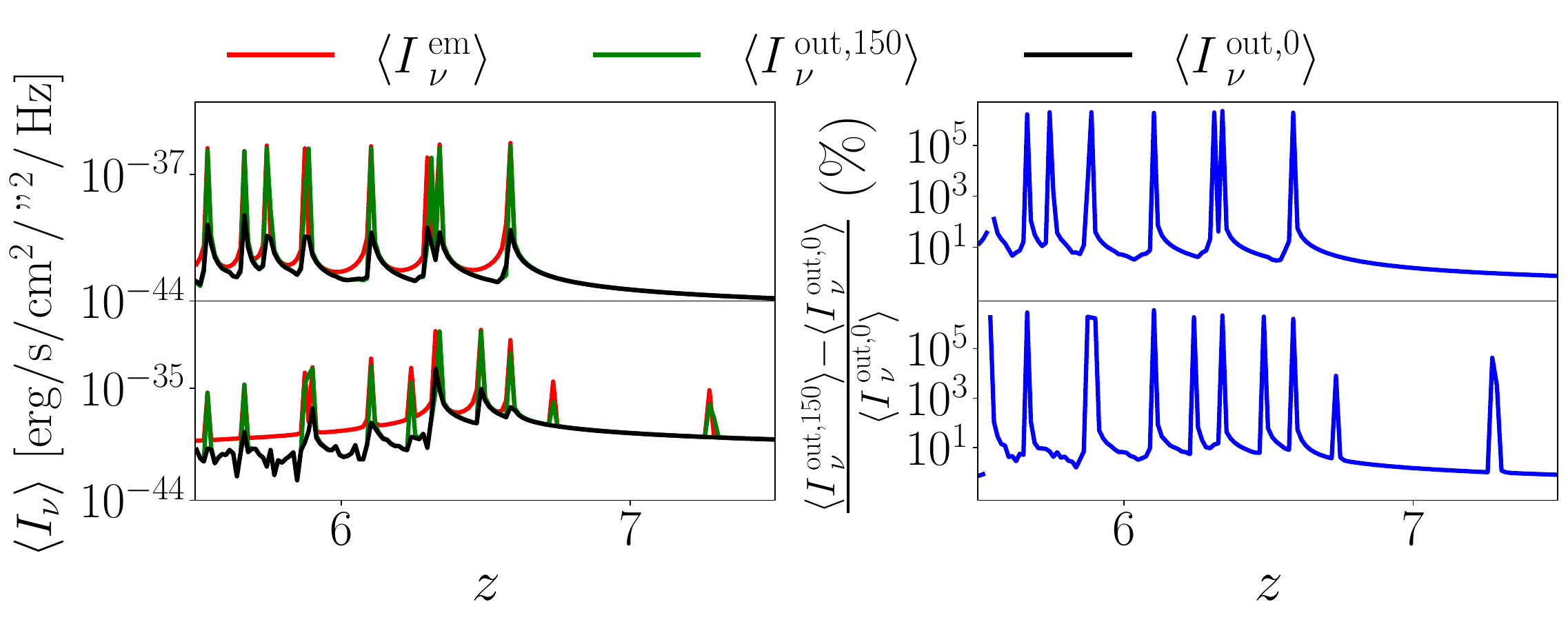}
    \caption{The spectral intensity from stars shown for two different sightlines (one in each row) for our outflow model when there is no velocity offset, shown in black, vs a velocity offset of $150\,\rm{km\,s}^{-1}$, shown in green. The pure spectral intensity is shown in the left column, and the fractional boost from including outflows is shown in the right column. There is significant enhancement in the model with boosted transmission due to galactic outflows. However, this is intended to illustrate the change in directly transmitted versus scattering-reprocessed signals.}
    \label{fig:compare-outflow-pixels}
\end{figure*}
\begin{figure}
    \centering
    \includegraphics[width=\linewidth]{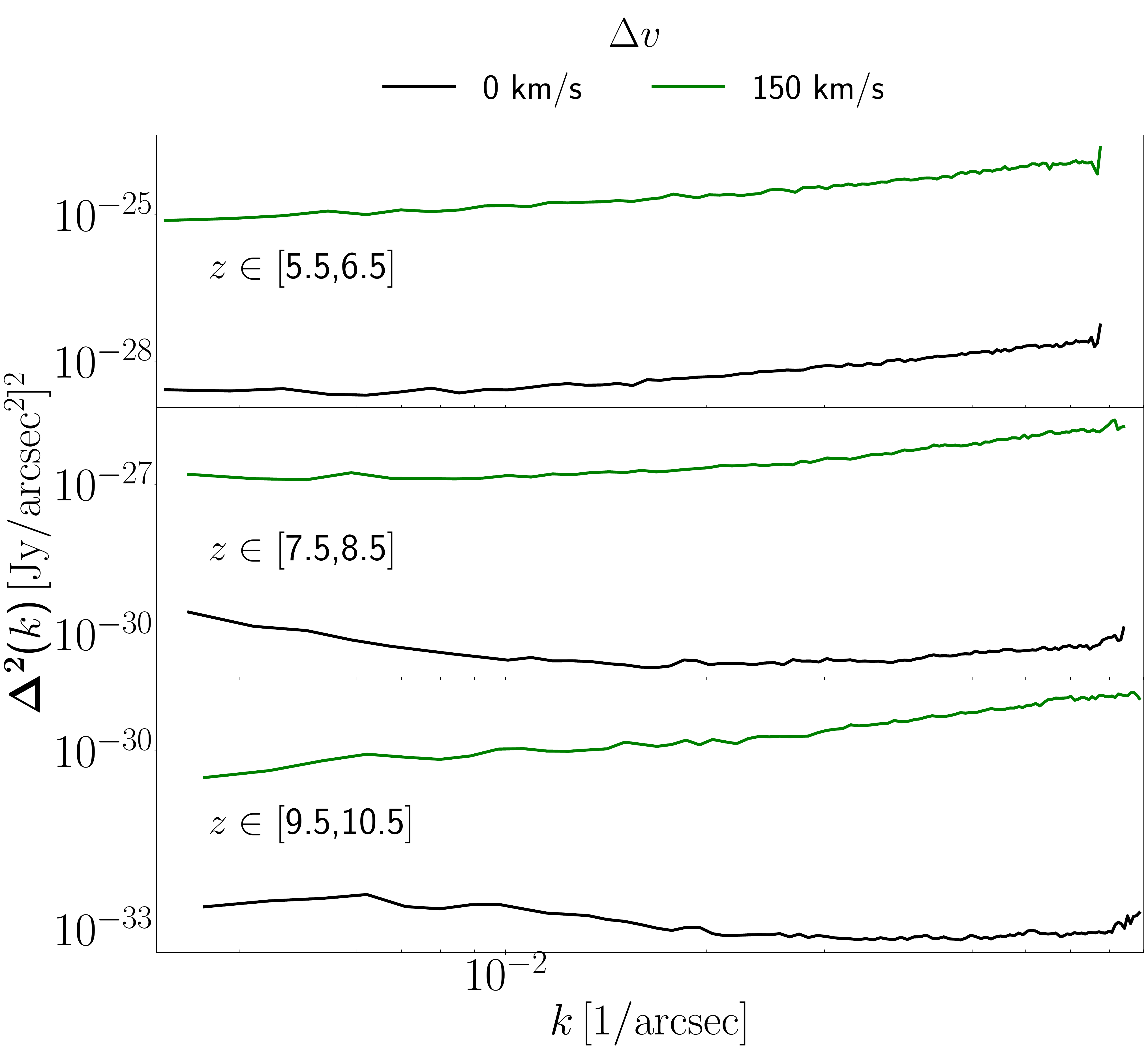}
    \caption{Resultant power spectra of the total emission channel maps for the two velocity offsets we simulate in our outflow comparison. We plot the total power spectrum when there is a redward velocity offset of $150\,\text{km\,s}^{-1}$ in green, and when there is no offset in black. Power spectra are plotted for different redshift ranges in each panel, with $z\in[5.5,6.5]$ shown in the top panel, $z\in[7.5,8.5]$ in the middle panel, and $z\in[9.5,10.5]$ in the bottom panel. At our lowest redshift band, we observe a boost of $\sim4\times$ in power when including the redward offset.}
    \label{fig:outflows-tot}
\end{figure}
In Figure~\ref{fig:power-spectra-emissionmech}, we plot 2D power spectra of the spectral intensity, now broken up by the constituent emission mechanisms. On the top panel, both emission and absorption are included in the radiative transfer, while the bottom panel shows spectra when just including emission sources (i.e., no absorption). Different columns show the spectra at different redshifts. It is worth noting that the emission from \HII regions tends to contribute significantly at the lowest redshift range [5.5,6.5] in the emission-only case (top right subplot), especially at the smallest scales, while being pushed down to the least dominant emission mechanism when including absorption (top left subplot). This difference could point towards the coincident self-shielding in massive galaxies where stellar emission is more dominant at these later redshifts. In the next sub-section, we test the validity of this theory.

\subsection{Exploring velocity offsets in the \HII region emission}
\label{subsec:outflows}

In this section, we briefly investigate the effects of sub-grid galactic outflows using a toy model. Specifically, we add a velocity offset to the unresolved emission from \HII regions to see if it helps address the drop in Ly$\alpha$ intensity that we observe from these sources at lower redshifts in our spectra. This offset is interpreted as boosting the directly transmitted component rather than the scattering-reprocessed one. We choose a nominal velocity offset $\Delta v=150\,\text{km\,s}^{-1}$, henceforth denoted as $\langle I\,_{\nu}^{\rm{out},150} \rangle$ to compare against the case where there is no velocity offset $\langle I\,_{\nu}^{\rm{out},0} \rangle$.

In Figure~\ref{fig:compare-outflow-pixels}, spectral intensities (left column) and their percent difference between the two velocity offsets (right column) from \HII region emission is shown for two representative pixels in our grid (one in each row). On the left panel, the standard result using a band-integrated Voigt profile is shown in red, with the outflow models shown in black and green for no velocity offset and a velocity offset $\Delta v=150\,\text{km\,s}^{-1}$ respectively. There is a clear boost in emission when including the velocity offset in both pixels, effectively recovering all of the emission-only signal (red curve). Although this is only shown for two sightlines, we can see the impact of self-absorption on the transmitted intensity, with the imposed redward offset allowing all of the emission to escape.

To definitively quantify the effect of including a velocity offset, which would also help in validating the observed suppression in the total signal when including absorption in the radiative transfer, we also plot the 2D power spectra of the total emission channel maps for the two velocity offsets in Figure~\ref{fig:outflows-tot}, with lines for $\Delta v = 0\,\text{km\,s}^{-1}$ and $150\,\text{km\,s}^{-1}$ in black and green respectively, and for different redshift ranges, with $z\in[5.5,6.5]$ plotted in the top panel, $z\in[7.5,8.5]$ in the middle panel, and  $z\in[9.5,10.5]$ in the bottom panel. It is interesting to note that there is a boost in the spectrum at all scales for the lowest redshift range, which is the range we saw to be most impacted by co-spatial absorption in our power spectra in the previous section. At $k\sim10^{-2}$\,--\,$10^{-1}$ arcsec$^{-1}$ in the lowest redshift band of $[5.5,6.5]$, we see an increase by a factor of almost four in the power spectrum signal. However, the increase in the power spectrum signal is also associated with an increase in Poisson noise introduced by point-like sources that are now offset from the absorption cells. It would be fruitful to explore the effects of including scattering on the resultant spectra, which we leave to a future paper on the topic.

\section{Discussion and Summary}
\label{sec:summary}
Our work represents a novel contribution and perspective on Ly$\alpha$ LIM modelling by using high-resolution, large-volume radiation-hydrodynamic simulations that self-consistently model the coupling between radiation and matter during the EoR. This allows us to account for the intricate processes affecting Ly$\alpha$ emission and its observability, ultimately enhancing our understanding of the early Universe and aiding in the interpretation of future observational data. Although the \thesanone box used in this analysis has a large volume for simulations of its kind, a box size of $L=95.5$\,cMpc at $z=5.5$ corresponds to an angular size of $\sim0.6$\,deg, meaning that SPHEREx's FOV of 3.5\,deg is $\sim 5\times$ larger. Future reionization simulations with much larger volumes will therefore improve large-scale LIM studies in the following ways: first, enhanced spatial statistics and extended simulated power spectra to larger angular scales; second, reduced periodic effects along the LoS by traversing larger volumes with on-the-fly light cones. These periodic effects are more visible when looking at the Ly$\alpha$ intensity along one sightline, as in Figure~\ref{fig:Lcum-Inu-3pix}, and are slightly more washed-out in global statistics, as in Figure~\ref{fig:total_intensity}.

The opportunity to probe IGM-scale radiative transfer at high resolution, as demonstrated by the channel maps in Figures~\ref{fig:channel_maps_inset_1024} and~\ref{fig:6-channel-maps}, proves to be useful even with the relatively low spatial resolution typical of LIM instruments. Our highest-resolution grid's pixel scale of $2.11''$ at $z=5.5$ is comparable to SPHEREx's pixel scale of $6.2''$, which is found to undersample its optical PSF's FWHM of $2.01''/\,$pixel at $\lambda=0.75\mu \rm{m}$ ($4.3''/\,$pixel when including aberrations and manufacturing constraints) \citep{Korngut_2018,Symons_2021}.

We found that the shape of the power spectrum is significantly impacted by the process of reionization, with the slope at smaller scales becoming steeper as the neutral fraction of the IGM decreases, supporting the findings of \citet{Visbal_2018} and \citet{AmVi2025}.  In works such as \citet{AmVi2025}, the Ly$\alpha$ LIM signal turns out to be comparably detectable by SPHEREx in comparison to our emission+absorption signal; however, the detectability of our emission-only power spectra, shown in Figure~\ref{fig:channel_maps_emissionmech}, along with the recovery of our emission-only \HII region signal when incorporating a velocity offset, shows the drastic effect of immediate self-absorption within the vicinity of overdense regions. The emergent spectra boost from the inclusion of outflows supports the conclusions of radiative transfer calculations in individual galaxies with transmission through the IGM \citep{Smith2019,Garel2021,Smith_2022}; however, subsequent IGM scattering in \citet{AmVi2025} seems to dull out these effects. By incorporating scattering self-consistently in the evolving IGM, we will be able to conclude whether the predicted power spectrum signal falls closer to the emission-only or the emission-and-absorption models. We leave the formation of these more definitive conclusions to future work.

We can thus summarize our conclusions as:
\begin{enumerate}[label=\Roman*.]   
    \item Employing a Voigt-profile treatment for absorption-only transmission substantially improves the agreement between our projections and observations.
    \item The median Ly$\alpha$ spectral intensity increases steeply for $z \lesssim 7$, converging towards the mean as the spatial resolution is decreased.
    \item A plateau in intensity as we approach $z \sim 5.5$ indicates the end of reionization, offering a signature for detecting the final stages with LIM techniques.
    \item Our emission-only signal is detectable by SPHEREx's surface brightness noise levels with reasonable assumptions, but our lower-limit emission and aborption signal is not.
    \item Power spectra for different redshift ranges of our LIM spectral cuboid show clear evolution with redshift, with the slope at smaller scales steepening with decreasing redshift.
    \item Including a redward offset in emission from \HII regions due to outflows leads to a boost in emission, especially at the smallest scales at lower redshifts (a factor of $\sim4\times$ in the power spectrum at a spatial resolution of $8.4''$).
\end{enumerate}
High-resolution simulations that accurately model the physics of the IGM are hence essential for accurately capturing diffuse emission during reionization, aligning with the primary goal of LIM studies to observe large-scale emission over cosmic time. This work focused on constructing a pipeline for generating Ly$\alpha$ LIM predictions from high-resolution renders of radiation hydrodynamic codes, but future work will refine this by incorporating additional observational factors (e.g., interference from line interlopers), facilitating direct comparisons with SPHEREx and other LIM instruments.

\section*{Acknowledgements}

% We thank the referee for constructive comments and suggestions which have improved the quality of this work.
AS acknowledges support through HST AR-17859, HST AR-17559, and JWST AR-08709.  RK acknowledges support of the Natural Sciences and Engineering Research Council of Canada (NSERC) through a Discovery Grant and a Discovery Launch Supplement (funding reference numbers RGPIN-2024-06222 and DGECR-2024-00144) and York University's Global Research Excellence Initiative. LH acknowledges support by the Simons Collaboration on ``Learning the Universe''.

% The Acknowledgements section is not numbered. Here you can thank helpful colleagues, acknowledge funding agencies, telescopes and facilities used etc. Try to keep it short.
% Remember to thank the referee.

%%%%%%%%%%%%%%%%%%%%%%%%%%%%%%%%%%%%%%%%%%%%%%%%%%
\section*{Data Availability}
Data products from \thesan are available online at \href{https://www.thesan-project.com}{www.thesan-project.com} as described in \citet{Garaldi2023}. Further data underlying this article will be shared on reasonable request to the corresponding author.

\newpage
\bibliographystyle{mnras}
\bibliography{biblio}
\newpage
\appendix

\section{COMPARISON OF SKEWERS VS CARTESIAN RENDERS}
\label{appendix:skewersvrenders}
As discussed in Section~\ref{sec:methods}, we employ 3D uniform cartesian renders in our exploration of the Ly$\alpha$ absorption and emission properties. In more detail, these renderings are mass-weighted projections of the original Voronoi mesh onto a coarser-grained uniform grid. The benefit of these renders is that they are much easier to store and manage than the other option: skewers that run through the original Voronoi mesh and preserve the raw data along the line of sight (i.e., no need to perform any averaging over Voronoi cells). The skewer format is preferable for certain applications due to possible discrepancies introduced by projecting the data onto a uniform grid, but the cartesian renders allow for better spatial sampling of the box, as well as better sampling in redshift, due to the computationally expensive process of producing the memory-intensive snapshot data and skewers in comparison.

\begin{figure}
    \centering
    \includegraphics[width=\linewidth]{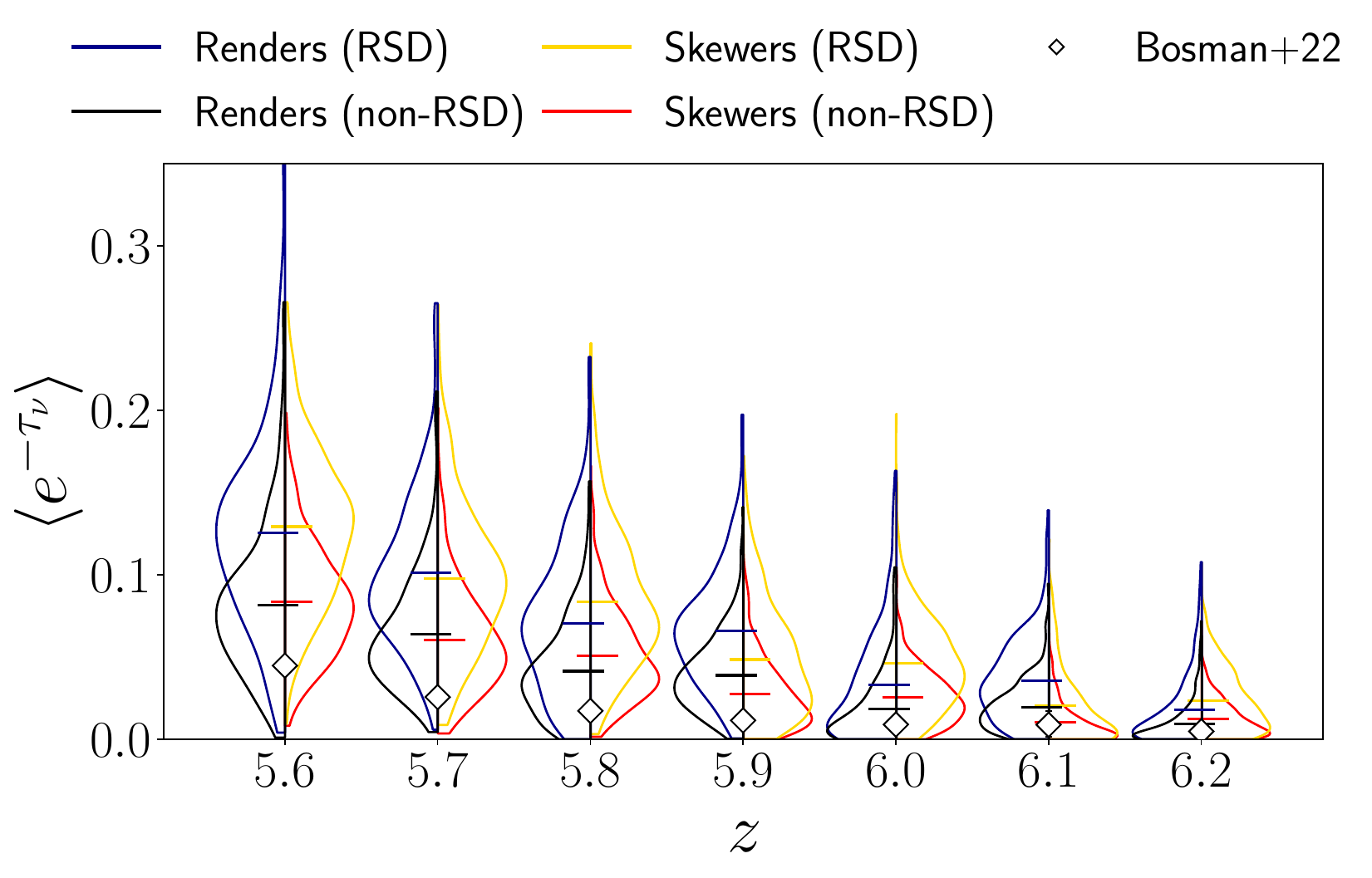}
    \caption{Violins of the Ly$\alpha$ transmission distribution for $2500$ random sightlines at different redshifts when taken from the cartesian renders, plotted in dark blue and black (RSD-corrected and non-corrected respectively) on the left side of each violin, vs. from the skewers, plotted in red and gold (RSD-corrected and non-corrected respectively) on the right side of each violin.}
    \label{fig:transmission_skewervrender}
\end{figure}

\begin{figure}
    \centering
    \includegraphics[width=\linewidth]{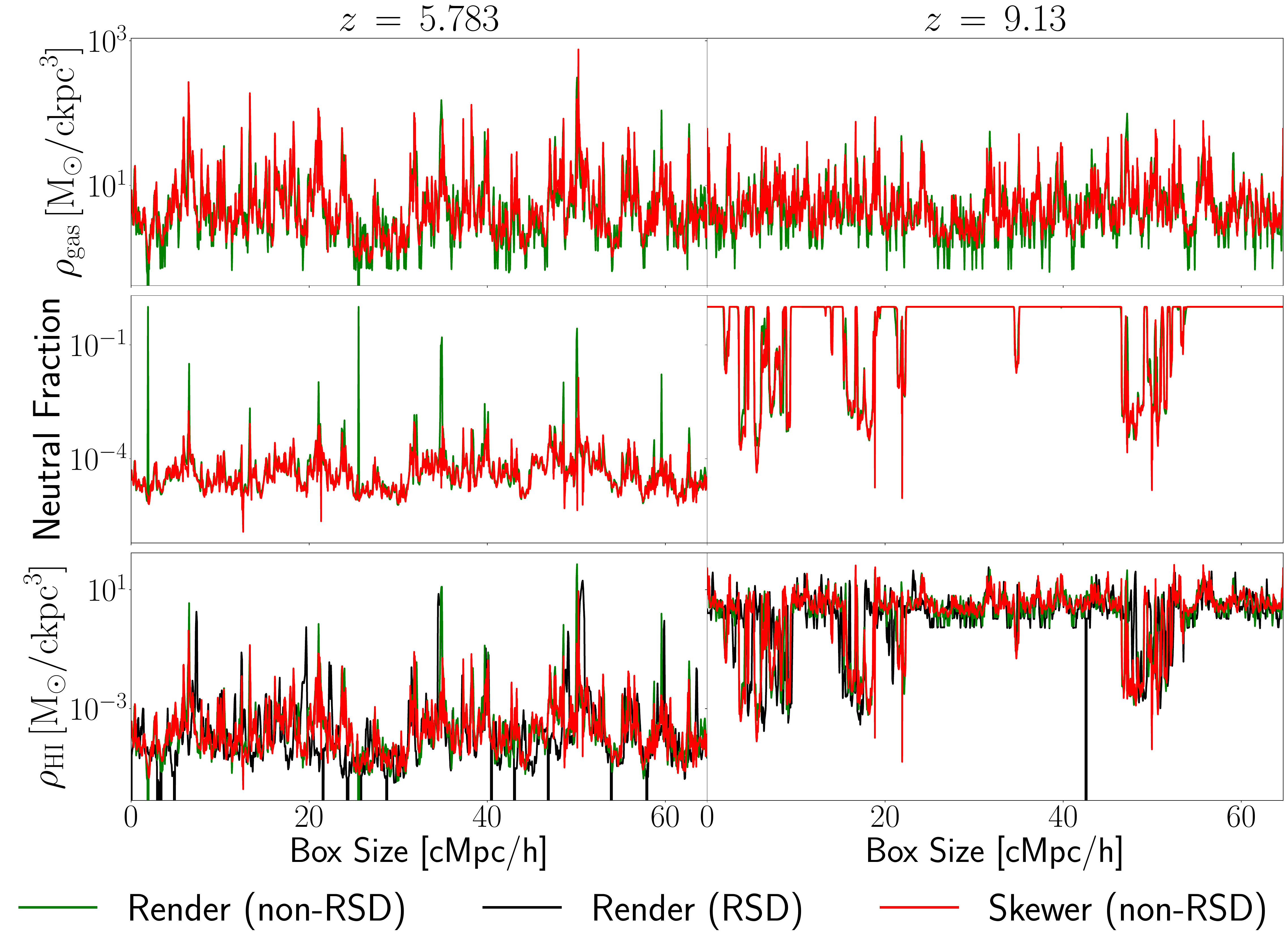}
    \caption{Gas mass density $\rho_{\rm{gas}}$, \HI Fraction, and neutral hydrogen density $\rho_{\HI}$ along the $z$ axis of the box at two different redshifts are plotted in each row respectively, with $z=5.783$ on the left panel, and $z=9.13$ on the right panel. The non-RSD corrected render data for one sightline is plotted in green, the RSD-corrected render data in black (only shown in the bottom panel), and the skewer data is plotted in red.}
    \label{fig:snapshot_skewervrender}
\end{figure}

\begin{figure}
    \centering
    \includegraphics[width=\linewidth]{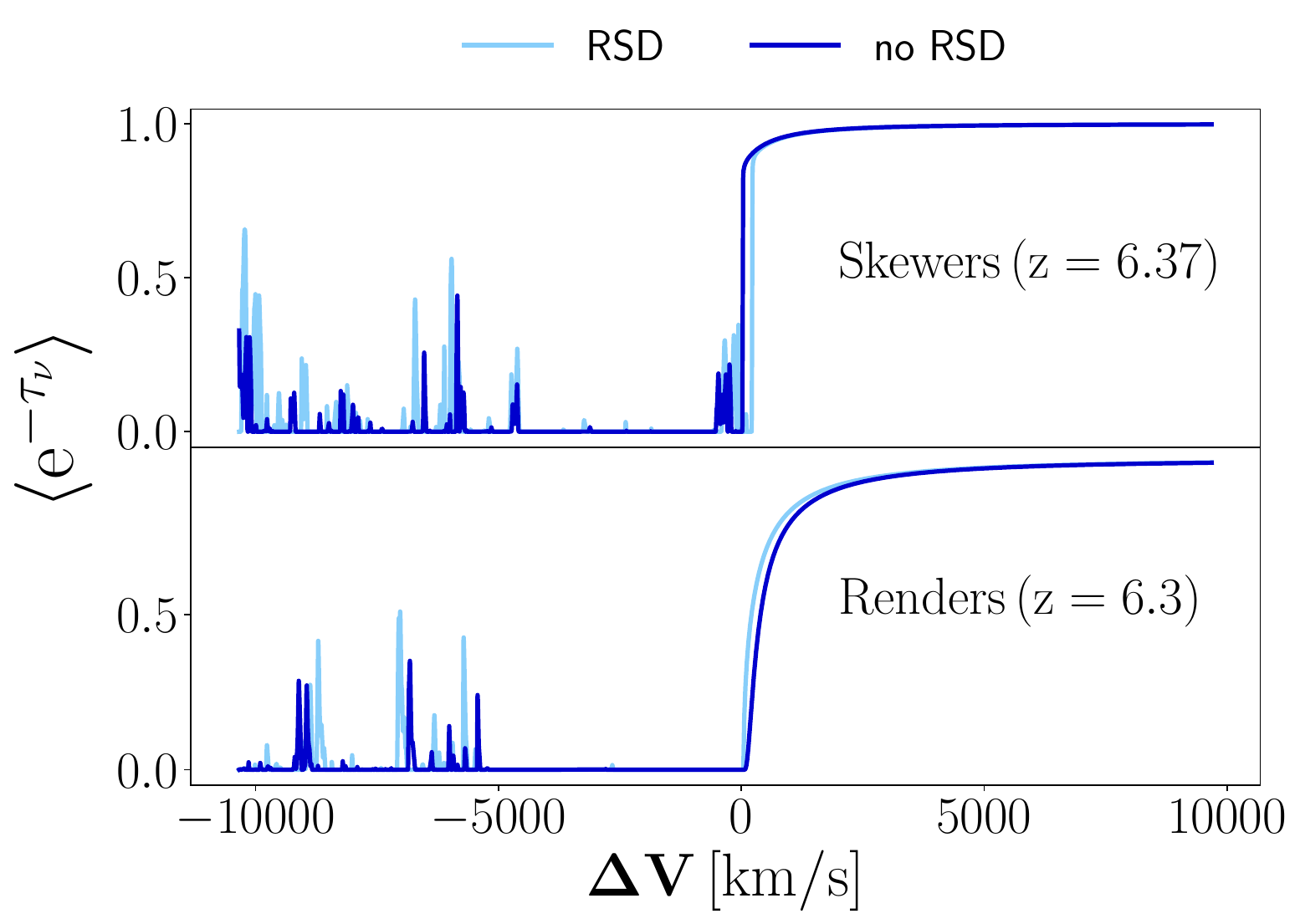}
    \caption{We plot the resulting transmission for a random sightline in each of the skewer (top panel) and render (bottom panel) lightcones when correcting for RSD (light blue) vs. not (dark blue). The chosen ource redshifts are from the random sample generated for our analysis. Note that the sightlines/redshifts for each of the skewer and render plots are unrelated, and the comparison is simply meant to show the consistent impact of including RSD corrections in each case, i.e., visibly decreased transmission in the case of no RSD.}
    \label{fig:compare-RSD-sightlines}
\end{figure}

To explore possible discrepancies first at a higher level, we employ a setup similar to that in Figure~\ref{fig:quasar-sightlines}, producing the transmission properties over a large number of sightlines in our light-cone both by using cartesian renders and skewers, and when incorporating the effects of Redshift Space Distortions (RSD) vs. not. This comparison is displayed in Figure~\ref{fig:transmission_skewervrender} in the form of violin plots, with the results from the render data plotted on the left side at each redshift, and the results from the skewers plotted on the right. Interestingly, incorporating RSD into the radiative tranfer leads to a much more dramatic shift than when using skewers as opposed to uniform cartesian renders. In other words, the higher spatial resolution introduced by the skewers does not play as important of a role as the resolved spectral shifts due to RSD. This result reassures us that employing cartesian renders for our purposes, hence decreasing memory usage and computational cost, is sufficient. In addition, it emphasizes the importance of including RSD effects in such analyses.

We also plot the raw gas mass density ($\rho_{\rm{gas}}$), neutral hydrogen density $\rho_{\HI}$ (note that this is RSD corrected for the renders but not the skewers), and neutral fraction along the LoS dimension for one pixel/skewer in the box in Figure~\ref{fig:snapshot_skewervrender} at two representative redshifts. Note that these data are not from the light-cones, but the original snapshots/renders to enable an apples-to-apples comparison between the two rendering methods\footnote{The snapshots and renders are saved at different cadences, so the light-cones constructed from each do not necessarily overlap due to there being more given wrap-arounds for the former.}. In the second row, we see that the render pixel's $\rho_{\HI}$ has sudden jumps and drops in comparison to the skewer; these jumps are especially prevalent at lower redshift (left column) due to the more frequent crossing of galaxies -- dominated by neutral gas by mass -- hence biasing the mass-weighted render data. The sudden drops in $\rho_{\HI}$ seem to follow the drops in gas density, and so are likely due to missing small-scale clumping corresponding to higher local densities in the Voronoi mesh. It is possible that employing volume weighting--rather than mass weighting--in the construction of light-cones could decrease the spikes seen in neutral fraction from coarse-grained data representations. We note that there is a significant difference introduced by the inclusion of RSD correction for both the renders and skewers, as was reflected in the transmission distributions we obtained in Figure~\ref{fig:transmission_skewervrender}. We can also see this on a sightline-by-sightline basis in Figure~\ref{fig:compare-RSD-sightlines}, where the top panel shows the effect of including RSD (in light blue) vs. not (in dark blue) for the skewers, and the bottom panel shows the same for the renders, each at random respective redshifts. Note that the one-to-one comparison can only be made between the two lines plotted in each row, and not between the skewer and the render spectra. We can see that in both cases, including RSD corrections leads to the enhancement of transmission, likely due to the displacement/effective ``smearing'' of neutral regions across cells.
\section{Comparison of integration methods}

\begin{figure}
    \centering
    \includegraphics[width=\columnwidth]{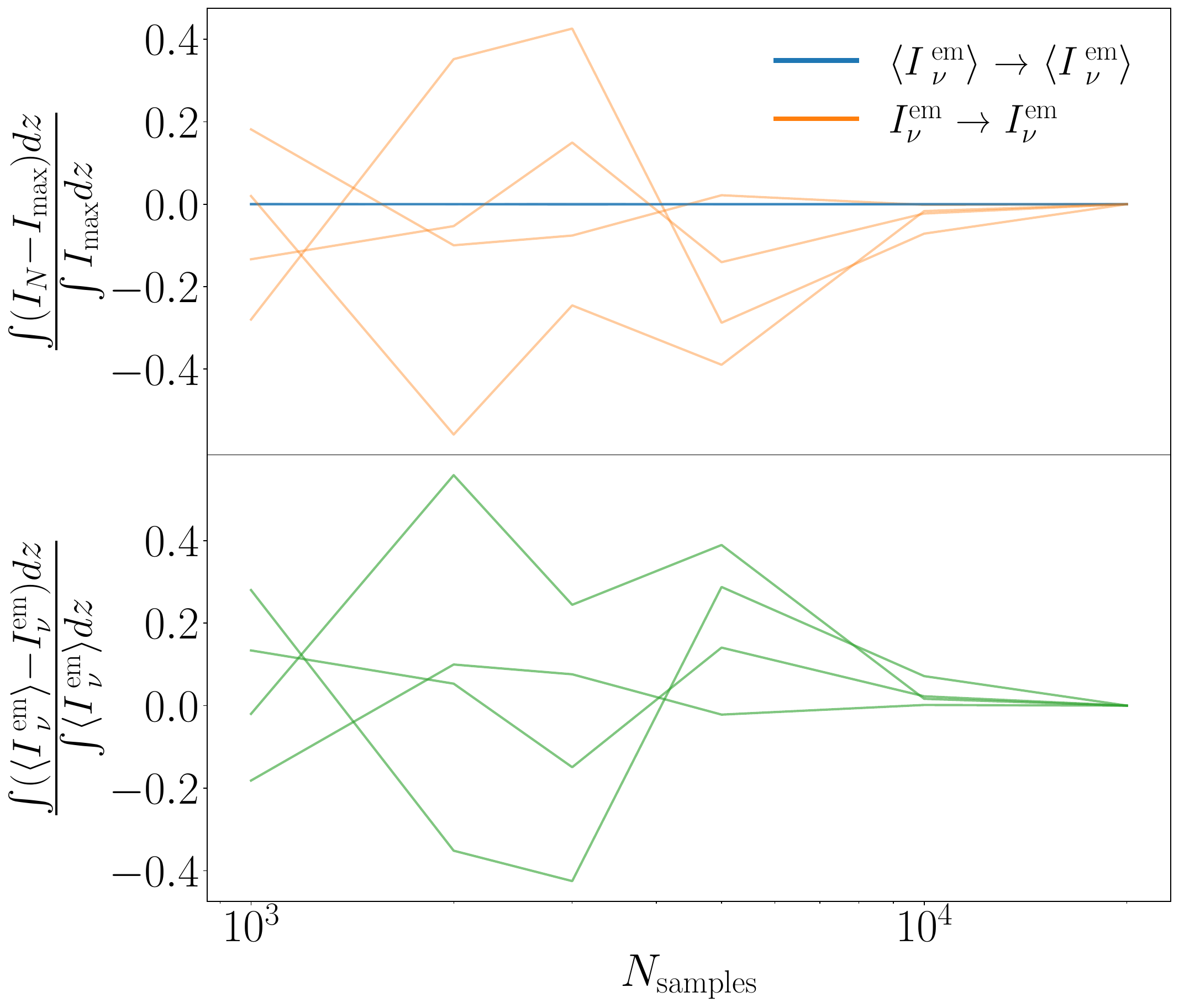}
    \caption{The top panel shows fractional residuals of the emission-only spectral intensities for different spectral samplings $N$ in comparison to that at the highest sampling rate $N_{\rm max}=2\times10^{4}$ for a few different sightlines, both when employing comb sampling (blue curves) vs band integration (orange curves) to test the convergence of each method as a function of spectral sample size. In the bottom panel, the fractional difference between the emission-only intensities produced by band integration and comb sampling methods are plotted as a function of spectral sample size for a few different sightlines.}
    \label{fig:compare_bandvcomb}
\end{figure}

In this paper, we derived a novel analytic solution for the band-integrated Voigt profile to improve the accuracy of our integration as compared to the baseline (comb sampling in frequency). In this section, we briefly explore the possible improvements in adopting this band integration method for $H(x)$; we will denote band integration as $\int_{\Delta x}H(x)$, and denote the comb sampling version as simply $H(x)$. 
% \ma{let's maybe show improvements for delta func too? those are obvious}
\begin{figure}
    \centering
    \includegraphics[width=\columnwidth]{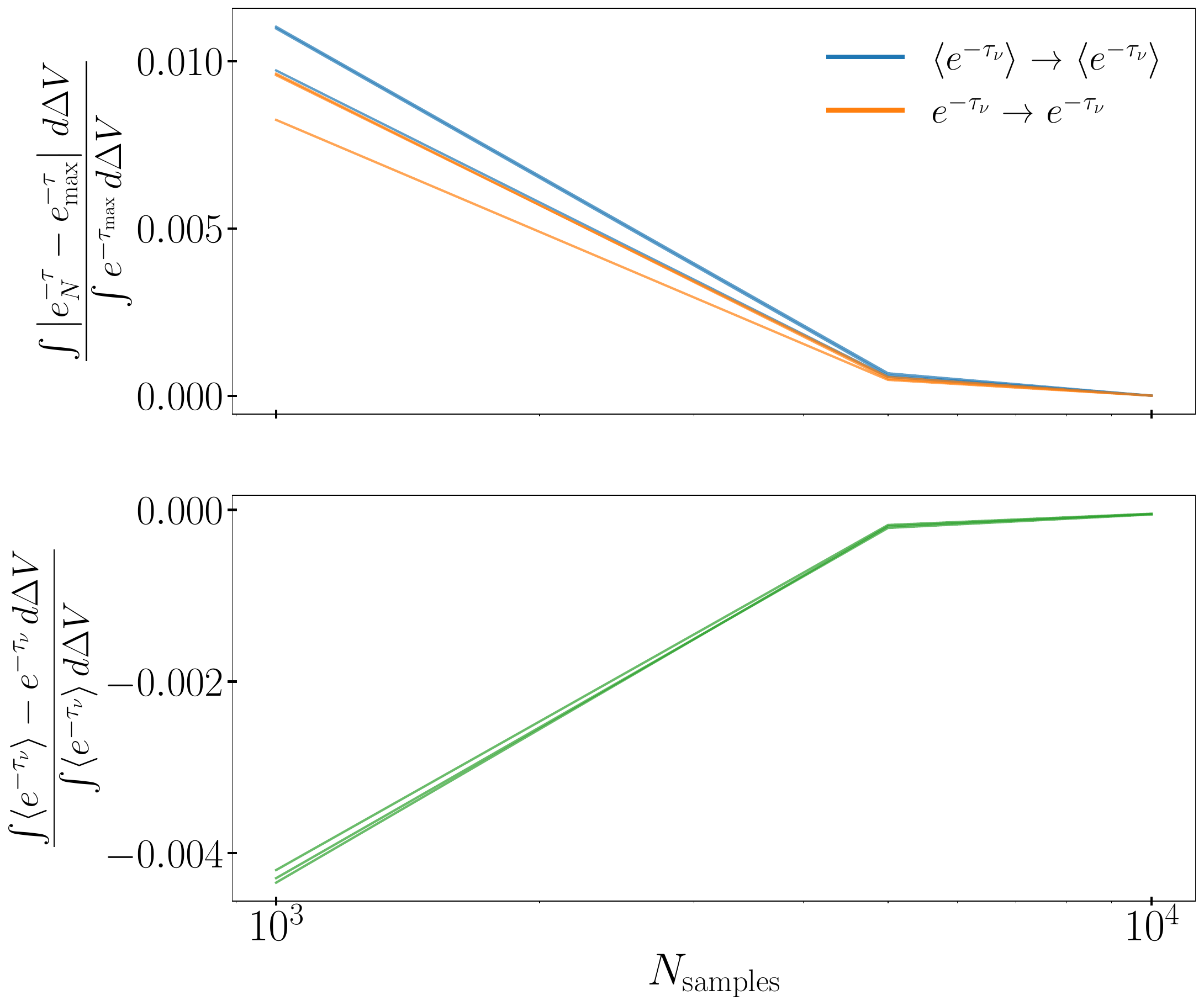}
    \caption{The top panel shows fractional residuals of the absorption-only transmission for different spectral sampling sizes $N$ in comparison to that at the highest sampling rate $N_{\rm max}=2\times10^{4}$ for a few different sightlines, both when employing comb sampling (blue curves) vs band integration (orange curves) to test the convergence of each method as a function of spectral sample size. In the bottom panel, the fractional difference between the absorption-only transmission produced by band integration and comb sampling methods is plotted.}
    \label{fig:compare_absonly}
\end{figure}

In the top panel of Figure~\ref{fig:compare_bandvcomb}, we show fractional residuals \[ \frac{\int (I_N -I_{\rm max})\,dz}{\int I_{\rm max}\,dz}\] of the emission-only spectral intensities for different spectral samplings $N$ in comparison to that of our highest sampling rate $I_{\rm max}$ (($N_{\rm max}=2\times10^{4}$ here); we show this metric for a few sightlines when employing comb sampling vs band integration to test the convergence of each method as a function of spectral sample size. The band-band residuals are shown as blue lines, whereas the comb-comb residuals are shown as orange lines; the band-band results are clearly superior, with the residuals for all sightlines sitting close to $0$ regardless of the adopted spectral sampling rate; in the comb sampling case, we can see that we start to reach convergence at $N=10^{4}$. In the bottom panel, we plot the fractional difference between the comb sampling and band-integration results for the same random sightlines, once again reaching convergence at a sample size $N=10^{4}$.
% In Figure~\ref{fig:compare_bandvcomb}, we plot the fractional residuals between the emission-only spectral intensities resulting from different sampling sizes for a single sightline; specifically, the more densely-sampled spectra (with some sample size $n=X$) are downsampled to a resolution of $10^{3}$, and we take the absolute difference between the two $I_{(n=X)}-I_{(n=5\times10^{3})}$. We can see clear suppression in the discrepancy between sample sizes when adopting band integration (right panel) versus comb sampling (left panel) of $H(x)$.

We similarly plot absolute residuals for absorption-only radiative transfer in Figure~\ref{fig:compare_absonly} for a few sightlines, with the top panel showing the band-band and comb-comb residuals (in comparison to our maximum sample size $N_{\rm max}=2\times10^{4}$) as a function of sample size $N$. We can see that the band integration and comb sampling methods are comparable for absorption-only analyses, quickly converging at a spectral sample size $N=5\times10^{3}$. Even at our smaller sample size of $10^{3}$, our residuals are minimal--at the sub percent level. In the bottom panel computing the fractional difference between comb sampling and band integration for a few sightlines, we similarly see very small residuals.

\section{Adopting an Analytic Form for Outflows}
\label{appendix:outflows}
\begin{figure}
    \centering
    \includegraphics[width=\columnwidth]{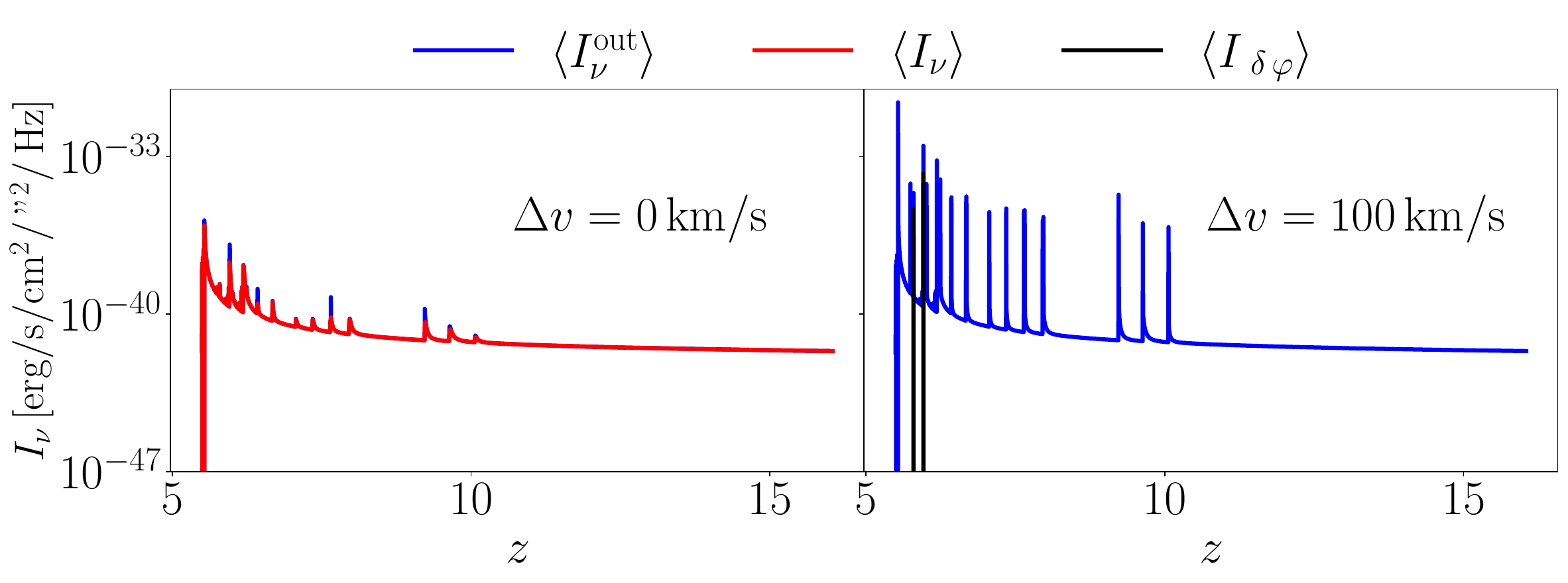}
    \caption{Spectral intensity from \HII Region emission (including absorption) for a representative pixel in our highest-resolution grid, with the left panel showing results when adopting no velocity offset, vs. the right panel when we adopt $\Delta v=100\,\text{km\,s}^{-1}$. All lines are for a spectral sample size $N=5000$. The result for the band-integrated $\langle I_\nu \rangle$ profile including self-absorption is plotted in red in the right panel, along with $\langle I\,_\nu^{\rm out} \rangle$ (not including self-absorption) in blue, and the $\langle I\,_{\delta\varphi} \rangle$ (i.e., Dirac delta emission and Voigt absorption) combination in black. The right panel shows results only for the latter two profiles as there is no analytic solution for outflows when we have $H(x)$ emission and absorption with self-absorption. There is very little transmission for \HII regions in the $\langle I\,_{\delta\varphi} \rangle$ case, rendering it unusable for our outflows analysis.}
    \label{fig:dv-combo}
\end{figure}
As was noted in an earlier section, incorporating a velocity offset into the unresolved \HII region emission would require a numerical solution if adopting Voigt profiles for both the emission and absorption. To reach an analytic solution, we explored a variety of combinations of functional forms for the emission and absorption. The first such combination that we attained an analytic solution for continued to employ a Voigt Profile for the absorption, but the frequency-dependent emission was instead represented by a Dirac delta function at line-center: $j_\nu = j_0 \,\delta(x - x_\text{out})$. Recall that $x_\text{out}$ is an optional redshifting term induced by outflows that is only given to photons originating from \HII regions. Solving Eq.~(\ref{eq:OG_RT}) with this model, we obtained the comb-sampling solution in Equation~\ref{eq:dv-comb-absem}, and the band-integrated in Equation~\ref{eq:dv-band-absem}.

We show the result of this formalism for a pixel in our grid in Figure~\ref{fig:dv-combo}, in the form of spectral intensity from \HII region emission. Despite reaching an analytic solution, we can see in the left panel that we are not able to replicate the spectral intensity produced by the Voigt profiles (in red and blue) and no velocity offset. This fact is true for \HII region emission specifically, which is concentrated along the LoS (increasing the likelihood of line skipping) and overwhelmed by self absorption. Including an offset of $150\,\text{km\,s}^{-1}$ (right panel) does seem to produce emission, but this model overall is not able to reliably produce our Ly$\alpha$ spectra as compared to the $H(x)$ model without  self absorption. For this reason, we resorted instead to the model that employs the band-integrated voigt profile for both emission and absorption, but does not include self-absorption to produce an analytic solution (Equation~\ref{eq:voigt-band-noSA}).

\section{Impact of spatial coarse graining on power spectra}
\begin{figure*}
    \centering
    \includegraphics[width=0.9\textwidth]{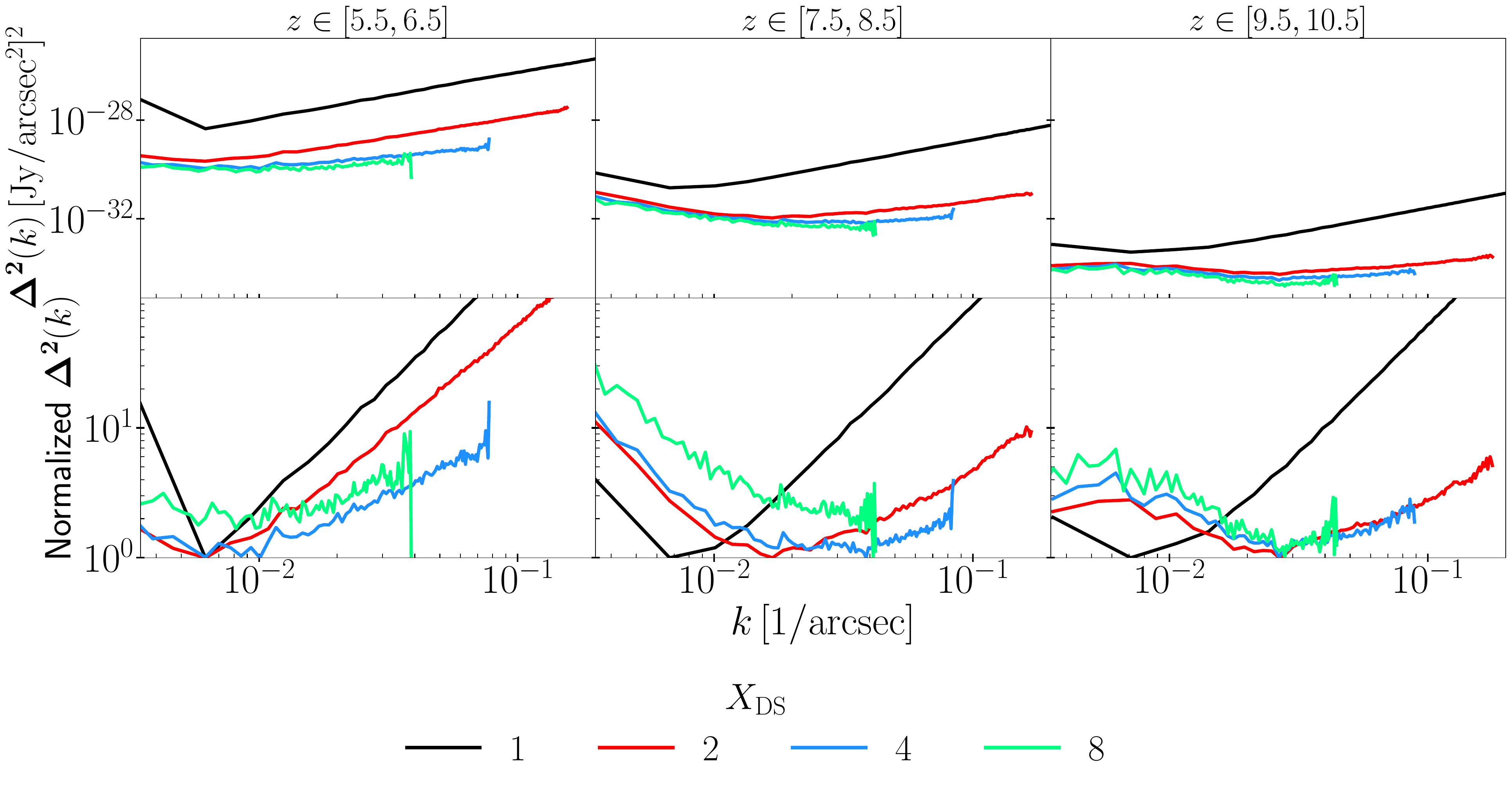}
    \caption{2D power spectra of channel maps of the total spectral intensity for different redshift ranges when employing a band-integrated Voigt profile for our emission and absorption; each redshift range occupies one column, with the top column showing the absolute spectra and the bottom column showing normalized spectra for different spatial downsampling factors $X_{\text{DS}}$ of our highest resolution grid. We show results from the highest resolution ($2.1''$) at $z=5.5$) in black, $X_{\text{DS}}=2$ ($4.2''$) in red,  $X_{\text{DS}}=4$ ($8.4''$), closest to SPHEREx's resolution) in blue, and $X_{\text{DS}}=8$ ($16.8''$) in green.}
    \label{fig:compare-PS-downsamplings}
\end{figure*}
In computing the radial power spectra of our channel maps, we find that in our current integration scheme, which includes emission and absorption only (i.e., no scattering), we find that we must employ spatial downsampling of our channel maps to mitigate the dominant shot noise component that appears when using the highest resolution $1024\times1024$ grid. In Figure~\ref{fig:compare-PS-downsamplings}, we show our power spectra for different redshift bands, with the top panel showing channel map power spectra for $z\in[5.5,6.5]$, the middle panel for $z\in[7.5,8.5]$, and the bottom panel for $z\in[9.5,10.5]$. In the left column, we show the absolute power spectra for different downsamplings of our spatial grid $X_{\text{DS}}$,and the right column shows the normalized power spectra. For context, these downsamplings of $\times 1$, $\times 2$, $\times 4$ and $\times 8$ correspond to pixel scales of  $2.1''$, $4.2''$, $8.4''$ and $16.8''$ respectively at $z=5.5$, and the pixel scale of SPHEREx is $\sim6''$. We can clearly see that computing the power spectra at the highest resolution produces a linear trend in the dimensionless power spectrum, which corresponds to a flat power spectrum (recall that $\Delta^{2}\propto k^{2}P(k)$ -- i.e., that of shot noise. We downsample our grid to smooth out point-like sources, revealing the underlying LIM power spectrum; it is worth noting that including scattering will likely smooth out the shot-noise component, and so this is simply an artifact of our relatively simpler model. In this paper, we chose to use the power spectra at a pixel scale of $8.4''$ at $z=5.5$ -- a downsampling factor of four -- as that is closest to the resolution of SPHEREx.

%%%%%%%%%%%%%%%%%%%%%%%%%%
% Don't change these lines
\bsp % typesetting comment
\label{lastpage}
\end{document}